\documentclass{article}

\usepackage{mathptmx}      \usepackage{graphicx} \usepackage{latexsym, amsfonts,amssymb,amsmath,amsthm} \usepackage{tikz, pgfplots, pgfplotstable}
\usepackage{exmath}  
\usepackage{xargs}
\usepackage{geometry}
\usepackage[ruled, vlined]{algorithm2e}
\usepackage{comment}
\usepackage{hyperref}
\usepackage{ulem}
\usepackage{enumitem}

\setlist[itemize]{leftmargin=*} \setlist[enumerate]{leftmargin=*}

\usepackage{etoolbox}
\usepackage{color}
\appto\ProcessRunnHead{\markboth{\truncate{220pt}{\authrun}\csuse{@author}}{\truncate{220pt}{\titrun}{\csuse{@title}}}}
\def\truncate#1#2{\ifdim#1<\wd#2\relax \rlap{\fboxsep=1pt\fboxrule1pt\color{red}\fbox{\hbox to#1{\strut\hss}}}\fi}

\usepackage[nocompress]{cite}
\theoremstyle{plain}
\newtheorem{theorem}{Theorem}[section]      
\newtheorem{lemma}[theorem]{Lemma}
\newtheorem{proposition}[theorem]{Proposition}
\newtheorem{assumption}[theorem]{Assumption}
\newtheorem{remark}[theorem]{Remark}

\newtheorem{text_comment}{Comment}{\bf}{\rm}

\usepackage{graphicx}
\usepackage{xcolor}
\hypersetup{  colorlinks,
    linkcolor={blue!50!black},
    citecolor={green!50!black},
    urlcolor={red!80!black}
}
\usepackage{url}
\hidedetails 

\RequirePackage{xargs}  \RequirePackage{xstring}

\newcommandx\filt[2][1=t, 2=F]{{\IfEq{#1}{set}{\nt{\mathbb #2}}{\nt{\mathcal{#2}_{{#1}}}}}}

\newcommand\pmeasure{\mathsf{P}}  \newcommand{\qmeasure}{\mathsf{Q}} \newcommand{\riskhor}{\mathnormal{H}}  \newcommand\loss[1][\riskhor]{\Lambda_{#1}}  \newcommand\pll{\varphi}  \newcommand{\lloss}{\lambda_{\pll}}
      \newcommand\sdedrift{\mu}  \newcommand\sdevol{\nt{\varsigma}}  
\newcommand\tol{\varepsilon}     

\newcommand{\thst}[1][none]{{\IfEq{#1}{none}{\theta^\star}{\theta_{#1}^\star}}}

\newcommand{\fmlmc}[1][none]{\IfEq{#1}{none}{\widehat F_{\epsilon, \sigma}^{\textnormal{(MLMC)}}}{\widehat F_{\epsilon_{#1}, \sigma_{#1}}^{\textnormal{(MLMC)}}}}
\newcommand{\pmlmc}[1][none]{\IfEq{#1}{none}{\widehat P_{\epsilon, \varsigma}^{\textnormal{(MLMC)}}}{\widehat P_{\epsilon_{#1}, \varsigma_{#1}}^{\textnormal{(MLMC)}}}}

  \newcommandx\MCell[2][1=M,2=\ell]{\hat E_{#1,#2}^{\p{\textnormal{MC}}}}  
    
\newcommand{\leb}{\mathcal L}
\newcommand\dl[1][\ell]{\Delta_{#1}}
\newcommand\Mvec[1][L]{\mathbf M_{#1}}
\newcommandx\MLMC[2][1=\Mvec, 2=L]{\hat E_{#1, #2}^{\p{\textnormal{MLMC}}}}

\newcommand\nestE[1]{U_{#1}}
\newcommand\appE[2]{\widehat U_{#1, #2}}
\newcommandx\antEc[3][1=0,2=\ell-1,3=1]{\widehat{U}_{{#3}, {#2}}^{\p{\textnormal{c}, #1}}}  \newcommandx\antEf[2][1=\ell,2=1]{\widehat{U}_{{#2}, {#1}}^{\p{\textnormal{f}}}}  \newcommand\appEub[2]{\widehat U_{#1, #2}^{\p{\textnormal{UB}}}}
\newcommandx\antEcub[3][1=0,2=\ell-1,3=1]{\widehat{U}_{{#3}, {#2}}^{\p{\textnormal{UB},\textnormal{c}, #1}}}  \newcommandx\antEfub[2][1=\ell,2=1]{\widehat{U}_{{#2}, {#1}}^{\p{\textnormal{UB}}}}  
\newcommandx\UMLMC[2][1=\Mvec, 2=L]{\widehat U_{0, #1, #2}^{\p{\textnormal{MLMC}}}}
\newcommandx\UMLMCT[2][1=\Mvec, 2=L]{\widehat U_{0, #1, #2}^{\p{\textnormal{MLMC}, T}}}

\newcommandx\appEml[2][1=\ell, 2=1]{\widehat{U}_{#2, #1}^{\p{\textnormal{ML}}}}  \newcommandx\antEcml[3][1=0,2=\ell-1,3=1]{\widehat{U}_{{#3}, {#2}}^{\p{\textnormal{ML}, \textnormal{c}, #1}}}  \newcommandx\antEfml[2][1=\ell,2=1]{\widehat{U}_{{#2}, {#1}}^{\p{\textnormal{ML}, \textnormal{f}}}}  \newcommandx\antEcmly[3][1=0,2=\ell-1,3=1]{\widehat{U}_{{#3}, {#2}}^{\p{\textnormal{ML}, \textnormal{c}, #1}}}  \newcommandx\antEfmly[2][1=\ell,2=1]{\widehat{U}_{{#2}, {#1}}^{\p{\textnormal{ML}, \textnormal{f}}}}  \newcommandx\Nl[2][1=0,2=1]{N_{#1}}  \newcommandx\Nlk[3][1=\ell,2=k,3=1]{\Nl[#1,#2][#3]}  

\newcommandx\antEfmlun[2][1=\ell,2=1]{\overline{U}_{{#2}, {#1}}^{\p{\textnormal{ML}, f}}}  \newcommand\ynest[2][none]{\IfEq{#1}{none}{
		Y_{#2}	
	}{
		Y_{#2,#1}	
	}
}
\newcommand\fnesty[1][none]{\IfEq{#1}{none}{f}{f_{#1}}}

\newcommandx{\Mlk}[2][1=\ell,2=k]{N_{#1,#2}}
\newcommandx\dljpath[2][1=j,2=\ell]{\Delta_{#1, #2}^{\p{\textnormal{path}, T}}f_{#1}}  \newcommand\yadv[4][none]{\IfEq{#1}{none}{\p{#2}_{#3,#4}}{\p{#2}_{#3,#4}^{\p{#1}}}}
\newcommand\yadvt[5][none]{\IfEq{#1}{none}{Y_{#4, #5}^{#3, #2}}{Y_{#4, #5}^{#3, #2; \p{#1}}}}

\newcommand{\ellmc}{{\vec{\ell}}}
\newcommand{\kmc}{{\vec{k}}}

\def\H#1{\I{#1 > 0}}

\newcommand\dirprior{\pi_{0,\ell}}

\newcommand\dhl[1][\ell]{\dl[\textnormal{ind}, #1]}

\newcommandx{\mlmcad}[2][1=\Mvec, 2=L]{\hat E_{#1, #2}^{\p{\textnormal{MLMC, ad}}}}

\newcommandx{\nvar}[3][1=none,2=none]{
	\IfEqCase{#3}{
		{x}{X\IfEq{#1}{none}{}{_{#1}}\IfEq{#2}{none}{}{^{({#2})}}}
		{y}{Y\IfEq{#1}{none}{}{_{#1}}\IfEq{#2}{none}{}{^{({#2})}}}
		{z}{Z\IfEq{#1}{none}{}{_{#1}}\IfEq{#2}{none}{}{^{({#2})}}}
		{sy}{y\IfEq{#1}{none}{}{_{#1}}\IfEq{#2}{none}{}{^{({#2})}}}
		{sz}{z\IfEq{#1}{none}{}{_{#1}}\IfEq{#2}{none}{}{^{({#2})}}}
		{d}{{V}\IfEq{#1}{none}{}{_{#1}}\IfEq{#2}{none}{}{^{({#2})}}}
	}
}
\newcommandx{\nvarbr}[4][1=none,2=none]{
	\IfEqCase{#4}{
		{x}{X\IfEq{#1}{none}{}{_{#1}}\IfEq{#2}{none}{^{\{#3\}}}{^{\{#3\}, ({#2})}}}
		{y}{Y\IfEq{#1}{none}{}{_{#1}}\IfEq{#2}{none}{^{\{#3\}}}{^{\{#3\}, ({#2})}}}
		{z}{Z\IfEq{#1}{none}{}{_{#1}}\IfEq{#2}{none}{^{\{#3\}}}{^{\{#3\}, ({#2}) }}}
		{sy}{Y\IfEq{#1}{none}{}{_{#1}}\IfEq{#2}{none}{^{\{#3\}}}{^{\{#3\}, ({#2})}}}
		{sz}{Z\IfEq{#1}{none}{}{_{#1}}\IfEq{#2}{none}{^{\{#3\}}}{^{\{#3\}, ({#2}) }}}
		{d}{V\IfEq{#1}{none}{}{_{#1}}\IfEq{#2}{none}{}{^{({#2})}}}
	}
} 
\newcommandx{\nvarcondbase}[6][1=none,2=none,3=none,4=none]{
 	\nvar[#1][#3]{#5}\p{\nvar[#2][#4]{#6}}
 }
\newcommandx{\nvarcond}[6][1=none,2=none,3=none,4=none]{
	\IfEq{#5}{d}{
		\nvar[#1][#3]{d}}{
		\IfEq{#2}{none}{
			{
				\nvarbr[#1][#3]{\nvar{#6}}{#5}
			}
		}{
			{
				\nvarbr[#1][#3]{\nvar[#2][#4]{#6}}{#5}
			}
		}
	}	
}
\newcommand{\rlev}[1]{
	\IfEqCase{#1}{{1}{\kappa_{1}}{0}{\kappa_{0}}{i}{\kappa_{i}}}
}

\DeclareMathOperator{\lgd}{\mathrm{LGD}}
\newcommand{\cva}[1]{\mathrm{CVA}_{#1}}
\newcommand{\adval}[1]{V_{#1}^{\p{\textnormal{CVA}, \pi}}}
\newcommand{\val}[1]{V_{#1}^{\pi}}
\newcommand{\adloss}[1][\riskhor]{\loss[#1]^{\p{\textnormal{CVA}, \pi}}}
\newcommand{\scs}{\sigma^{\p{\textnormal{cs}}}} \newcommand{\ct}[1]{c_{#1}}
\newcommand{\wcs}[1]{W_{#1}^{\p{\textnormal{cs}, \pmeasure}}}
\newcommand{\defc}[1]{{\tau_{\ct{#1}}}}
\newcommand{\srhor}[1][\riskhor]{
	\IfEq{#1}{0}{S_{#1}}{S_{#1}^{\pmeasure}}
	}
\newcommand{\defcv}{\Psi}
\newcommandx{\srt}[3][2=none,1=\riskhor]{S_{#3}^{{#1, \IfEq{#2}{none}{\srhor[#1]}{#2}}}}
\newcommand{\srtt}[3][\riskhor]{S_{#2}^{{#3, \srhor[#1]}}}
\newcommand{\fcva}[1]{f\p[\big]{{\nestE{1}}\p{\srt[#1]{\defc{#1}}, \defc{#1}}}}
\newcommand{\fcontrolvar}{f\p[\big]{{\nestE{1}}\p{\srt[\riskhor]{\defc{0}}, \defc{0}}}}
\newcommand{\fcvas}[1]{f\p[\big]{{\nestE{1}}\p{\srt[#1]{\tau}, \tau}}}
\newcommand{\srtpm}[3][0]{S_{#2}^{{#1, \srhor[#1], #3}}}
\newcommand{\fcvaspm}[2]{f\p[\big]{{\nestE{1}}\p{\srtpm[#1]{\tau}{#2}, \tau}}}
\newcommand{\fcontrolvars}{f\p[\big]{{\nestE{1}}\p{\srt[\riskhor]{\tau}, \tau}}}
\newcommand\lcva{\Phi}
\newcommand{\fb}{\bar f}
\newcommand{\volcva}{\varsigma}
\newcommand{\Mmell}[1][\ell]{N_{0,#1}}
\newcommand{\refp}{R}
\newcommand{\kbe}{\nt{A}}
\newcommand{\jbe}{\nt{B}}
\newcommand{\ncp}{{N_{\textnormal{cp}}}}

\pgfplotsset{compat=newest}

\pgfplotsset{width = 5.5cm,
			 every axis plot post/.append style = {line width = 1pt,
		 		every mark/.append style = {solid}}}
	 		
\newcommand\opac{0.6}  \newcommand{\nt}[1]{#1}
\newcommand{\ntg}[1]{#1}
\definecolor{revblue}{HTML}{2A52BE}
\newcommand{\st}[1]{#1}  \newcommand{\final}[1]{#1}
\newcommand\dpz{\final{V}}
\begin{document}

\title{Efficient Risk Estimation for the Credit Valuation Adjustment\thanks{The authors wish to express thanks for helpful comments \ntg{from two anonymous reviewers} on an earlier draft of this paper. JS was supported by the EPSRC Centre for Doctoral Training in Mathematical Modelling, Analysis and Computation (MAC-MIGS) funded by the UK Engineering and Physical Sciences Research Council (grant EP/S023291/1), Heriot-Watt University and the University of Edinburgh. 
}}
\author{
	Michael B. Giles\thanks{Mathematical Institute, University of Oxford, Oxford, OX2 6GG, UK.}
	\and Abdul-Lateef Haji-Ali\thanks{School of Mathematical and Computer Sciences, Heriot-Watt University, Edinburgh, EH14 4AS, UK and Maxwell Institute for Mathematical Sciences, Bayes Centre, 47 Potterrow, Edinburgh, EH8 9BT, UK}
	\and Jonathan Spence\footnotemark[3]\mbox{\,} \thanks{Corresponding author: \texttt{jonathan.spence@hw.ac.uk}}
}
	
\date{\today}

\maketitle
\begin{abstract}
The valuation of over-the-counter derivatives is subject to a series of valuation adjustments known as xVA, which pose additional risks for financial institutions. Associated risk measures, such as the value-at-risk of an underlying valuation adjustment, play an important role in managing these risks. Monte Carlo methods are often regarded as inefficient for computing such measures. As an example, we consider the value-at-risk of the Credit Valuation Adjustment (CVA-VaR), which can be expressed using a triple nested expectation. Traditional Monte Carlo methods are often inefficient at handling several nested expectations. Utilising recent developments in multilevel nested simulation for probabilities, we construct a hierarchical estimator of the CVA-VaR which reduces the computational complexity by several orders of magnitude compared to standard Monte Carlo.
%
%
\end{abstract} 
\section{Introduction}
Financial derivatives traded on over-the-counter markets are subject to a series of valuation adjustments, collectively known as \nt{x-}valuation adjustments (\nt{xVA}), reflecting the value of external risk factors. \final{A comprehensive summary of xVA can be found in the books of Gregory \cite{Gregory2020}, Green \cite{GreenAndrew2015XCFa} and Brigo et al.. \cite{BrigoDamiano2013Ccrc}.} The presence of \nt{xVA} components in valuation formulae pose an additional source of risk to financial organisations: Fluctuations in external risk factors can adversely affect the adjusted value of a financial portfolio. A common feature of \nt{xVA} calculations \nt{and associated risk measures} is a recursive or nested dependency on market factors, which \nt{poses} difficulties for numerical computation. This paper \nt{develops} efficient hierarchical estimators for such problems, with an application presented for a risk estimate  involving the credit valuation adjustment (CVA), \nt{which} accounts for the possibility of a counterparty defaulting and failing to meet future payoffs. To mitigate the possibility of fluctuations in \nt{the CVA} resulting in significant losses to a portfolio holder, the Basel Committee on Banking Supervision \cite{BaselIII} proposed a capital charge \final{to protect against such losses. Brigo et al.. \cite{BrigoDamiano2013Ccrc} and Pykhtin \cite{Pykhtin2012} relate this capital charge to a} \nt{CVA} value-at-risk formula over a short risk horizon \(\riskhor\ll 1\). At a \nt{sufficiently high} confidence level \nt{\(\pll\in\p{0,1}\)},
the (unilateral) CVA capital charge owed to a single counterparty can be expressed as the value \(\lloss\in \rset\) solving \cite{BrigoDamiano2013Ccrc}
\begin{equation}
	\label{eqn:cva_master}
	\begin{aligned}
		\pll &= \prob[\bigg]{\frac{\cva{\riskhor}}{B_\riskhor} - \cva{0} > \lloss},\\
		\cva{t} &=B_t\E[\Big]^{\qmeasure}{\I{t\le\tau\le T}\lgd\ntg{B_\tau^{-1}}\max\br[\big]{\E[\big]^{\qmeasure}{\pi_{\tau,T}\given\filt[\tau][G]}, 0} \given \filt[t][G]},
	\end{aligned}
\end{equation}
where \nt{\(\chi_A\) denotes the characteristic function of the event \(A\), taking value 1 if \(A\) is true and 0 otherwise. \st{For a given probability space \(\p{\Omega, \mathcal F, \qmeasure}\), we adopt the notation \(\E^\qmeasure{f}\defeq \int_\Omega f\text{d}\qmeasure\) to explicitly denote integration with respect to the measure \(\qmeasure\). \final{A derivation of $\cva{t}$, denoting the CVA priced at time $t$, can be found in Ahlberg \cite{Ahlberg2013CreditVA}.} The} remaining terms appearing in the above formula are defined as follows}: \st{The loss given default, denoted \(\lgd\), represents the (potentially random) proportion of future cash-flows which are lost if a counterparty defaults at time \(\tau\)}. \st{Real-valued} stochastic future payoffs \ntg{accrued} from default time \(\tau\) to maturity \(T\), \ntg{discounted to \(\tau\),} are denoted by \(\pi_{\tau,T}\) \ntg{and} assumed to be determined by \nt{\(\rset^d\)-valued assets \(S = \p{S_t}_{t\ge 0}\)} which have dynamics modelled through an SDE of the form
\begin{equation}\label{eqn:cva_sde}
\text{d}S_t = \sdedrift\p{t, S_t}\text{d}t + \sdevol\p{t, S_t}\text{d}W_t,
\end{equation}
under a physical market measure, denoted \(\pmeasure\), with drift \(\sdedrift:[0, T]\times \rset^d\to \rset^d\), volatility \(\sdevol:[0,T]\times \rset^d\to \rset^{d\times d^\prime}\) and where \nt{\(W = \p{W_t}_{t\ge 0}\)} is a \(d^\prime\)-dimensional Brownian motion adapted to a market filtration \(\filt[set] \defeq \p{\filt}_{t\ge0}\). \st{The numeraire, \((B_t)_{t\ge0}\), representing the value of a risk-free asset is a positive-valued stochastic process adapted to \(\filt[set]\), with \(B_0 = 1\). A future payoff \(\Pi_T\) at time \(T\), discounted to time \(t\le T\), is hence calculated as \(\pi_{t, T} = B_{t}B_{T}^{-1}\Pi_T\).}  We assume the existence of a risk-neutral measure \(\qmeasure\), which is used for arbitrage-free pricing of derivatives, such that the discounted assets \st{\((S_t/B_t)_{t\ge0}\) form a \(\qmeasure\)-martingale adapted to \(\filt[set]\)}. \st{Since the capital charge \(\lloss\) (correspondingly, the confidence level \(\pll\)) in  \eqref{eqn:cva_master} is a risk measure, it is computed under \(\pmeasure\) to reflect the true (risky) dynamics of the market}. \st{In contrast,} \(\cva{t}\) represents a valuation adjustment, and is priced under \(\qmeasure\). The random future time of counterparty default, \(\tau\), is a stopping time with respect to an
extended filtration \nt{\(\filt[set][G] \defeq \p{\filt[t][G]}_{t\ge 0}\) such that \(\filt[t][G]\supset \filt\) for all \(t\ge 0\)}, incorporating credit risk factors not present in the market filtration \(\filt[set]\). \st{Contextually, \(\lgd\) is assumed to be a \(\filt[\tau][G]\)-measurable random variable taking values in \(\p{0,1}\).} 
The term \ntg{\(\max\br{\E^{\qmeasure}{\pi_{\tau,T}\given\filt[\tau][G]}, 0}\)} is the  positive part of the (discounted) risk-neutral portfolio valuation at the default time \(\tau\) and is referred to as the exposure at default, representing the amount the portfolio holder stands to lose at the time of counterparty default. Only the positive part of the portfolio value is considered in the CVA formulae since the portfolio holder must still typically repay any negative values, for example to creditors, in the event of counterparty default. 

Due, for example, to complex market dynamics \eqref{eqn:cva_sde}, it is typically not possible to sample the assets \(S\), respectively the payoffs \(\pi_{\tau, T}\), exactly. Often, accurate approximations of the term \(\pll\) \eqref{eqn:cva_master} are obtained using \nt{expensive} time-discretisations to approximate the process \eqref{eqn:cva_sde}. Furthermore, due to approximation of the repeatedly nested expectations defining \(\cva{\riskhor}, \cva{0}\) and \(\pll\), the compounding cost of combining time-discrete approximations of the SDE \eqref{eqn:cva_sde} within nested averages leads to impractically large costs involved in approximating \eqref{eqn:cva_master}. 

Motivated by the problem of approximating the CVA capital charge \eqref{eqn:cva_master}, this paper considers efficient methods to evaluate the distribution of random variables which can be expressed as a sequence of nested expectations. Abstractly, for \(\pll\in\p{0,1}\) and \(\lloss\in\rset\), we consider problems taking the form
\begin{equation}\label{eqn:cva_general_nested_prob}
	\begin{aligned}
		\pll &= \prob[\big]{{\nestE{0}} > \lloss}\\
		\st{{\nestE{0}}} &= \E[\Big]{f\p[\big]{\st{{\nestE{1}}}}\given \nvar{z}}\\
		\st{{\nestE{1}}} &= \E{\nvar{x}\given \nvar{y}},
	\end{aligned}
\end{equation}
where \(f:\rset\to\rset\) is Lipschitz continuous, but not necessarily everywhere differentiable, and \(\nvar{x},\nvar{y}\) \st{and} \(\nvar{z}\) are random variables, with \(\nvar{x}\) taking values in \(\rset\) \st{and depending on} \(\nvar{y}\) \st{and, in turn,} \(\nvar{z}\), \st{both taking} values in \(\rset^d\). \st{Crucially, the method proposed in this paper does not assume that the variables \(\nvar{x}\), \(\nvar{y}\), \(\nvar{z}\), \({\nestE{0}}\) and \({\nestE{1}}\) can be sampled \st{exactly}, but instead} through a hierarchy of increasingly accurate approximations. Letting \(\ct{\riskhor}\) be an \(\rset^{d_c}\) random variable denoting the value at the horizon \(\riskhor\) of all relevant \(\filt[\riskhor][G]\)-measurable credit risk factors, influencing the default times \(\tau\), the CVA problem \eqref{eqn:cva_master} can be expressed in the form \eqref{eqn:cva_general_nested_prob}, \st{where}
\begin{equation}\label{eqn:cva_vars}
	\begin{aligned}
		\nvar{z} &= \p{S_{\riskhor}, \ct{\riskhor}}, \\
		\nvar{y} &= \p{S_\tau, \tau}, \\
		\nvar{x} &= \ntg{\lgd}\I{\riskhor\le\tau\le T}\ntg{B_\tau^{-1}}\pi_{\tau,T},
	\end{aligned}
\end{equation}
each \st{depend} on the solution to an \st{It\^o} SDE of the form \eqref{eqn:cva_sde} \st{which can be approximately sampled through an appropriate numerical method}. \st{Linking to \eqref{eqn:cva_master}, the random variable \(\nvar{z}\) represents risk factors and is sampled under \(\pmeasure\), whereas \(X\) and \(Y\) appear within the pricing of \(\cva{\riskhor}\) and are sampled under \(\qmeasure\).} The connection between the problem \eqref{eqn:cva_master} and the general equation \eqref{eqn:cva_general_nested_prob} is discussed further in Section\nobreakspace \ref {sec:cva_nest_exp}. \nt{\st{To simplify notation,} when considering the general problem \eqref{eqn:cva_general_nested_prob}, we assume each expectation and probability is computed with respect to a single, arbitrary measure, \(\pmeasure\). Alternatively, when considering the CVA problem \eqref{eqn:cva_master}, we explicitly track the dependency of each term on the physical measure, \(\pmeasure\), and the risk-neutral measure, \(\qmeasure\).}
We focus primarily on the problem of approximating the probability of large loss \(\pll\) given a fixed \(\lloss\). 
Such methods can then be \final{used within} stochastic root finding techniques to find the quantile \(\lloss\) corresponding to a fixed confidence level \(\pll\).

\nt{Monte Carlo estimators to approximate probabilities containing a single nested expectation of the form \(\prob{\E{X\given Y}>\lloss}\) \final{are considered by Gordy and Juneja \cite{Gordy:2010} and Brodie et al.. \cite{Brodie:2011}. Multilevel extensions to these estimators are further considered by Giles and Haji-Ali in \cite{GilesHajiAli:2018,GilesHajiAli:2019sampling,hajiali2021adaptive}.} Stochastic approximation methods to compute the corresponding value-at-risk \(\lloss\)  for such problems are considered \final{by Frikha et al.. \cite{Frikha2016,cfl:2023,bfp09,cflp:2023,adsa24} and by Dereich and M\"uller-Gronbach \cite{Dereich2019,Dereich2021}.} An application of \st{multilevel Monte Carlo (MLMC)} methods to problems defined by arbitrarily many nested Lipschitz transformations of conditional expectations is considered \final{by Wang et al.. in} \cite{syed2023optimal,zhou2022} for problems arising in optimal control and non-European option pricing. Within the CVA framework, multilevel methods have previously been employed \final{by Hofer and Karlsson} in \cite{hk17} to the problem of calibrating CVA model parameters to observed market data.}
The key contributions of this paper are as follows:
\begin{itemize}
	\item A novel combination of \st{MLMC} methods for nested expectations and probabilities, \final{extending the works} \cite{GilesHajiAli:2018,hajiali2021adaptive}, to compute \(\pll\) in \eqref{eqn:cva_general_nested_prob} is analysed in Section\nobreakspace \ref {sec:cva_mlmc}.  Under general assumptions on the underlying variables, Theorem\nobreakspace \ref {thm:ad_mlmc_hierarchy} proves that the proposed estimator approximates \(\pll\) with root mean square error \(\tol\) with order \(\tol^{-2-\delta}\) computational cost, for any \(\delta>0\). This is contrasted with an  order \(\tol^{-5}\) computational cost using \ntg{standard} Monte Carlo techniques for the same problem.
	\item Section\nobreakspace \ref {sec:cva_cva} discusses an implementation of the hierarchical estimator for the CVA risk problem \eqref{eqn:cva_master}. Section\nobreakspace \ref {sec:cva_cva_capital} leverages variance reduction techniques from the literature to further improve the efficiency of the estimator. Extensions of the model to include credit risk mitigation factors such as the posting of collateral and to portfolios with more than a single counterparty are discussed in Sections\nobreakspace \ref {sec:collateralisation} and\nobreakspace  \ref {sec:cva_multiple_counterparties}.   
	\item A numerical study to support theoretical results is conducted in Section\nobreakspace \ref {sec:cva_num}.
\end{itemize}

\section{Nested Multilevel Monte Carlo Estimator}\label{sec:cva_mlmc}
In this section we consider the construction of efficient MLMC estimators for general nested risk calculations. We begin by outlining key assumptions on (approximations) of the underlying variables in Section\nobreakspace \ref {sec:cva_setup_assumptions}, before considering a single-level Monte Carlo estimator in Section\nobreakspace \ref {sec:cva_hierarchical_mc} and multilevel Monte Carlo estimator in Section\nobreakspace \ref {sec:cva_unbiased_mlmc}.

\subsection{Setup and Notation}
\label{def:cond_rv}
{
Throughout this section, we consider the general problem \eqref{eqn:cva_general_nested_prob} defined for random variables \(\nvar{x}:\Omega\to\rset\) and \( \nvar{y},\ \nvar{z}:\Omega\to \rset^d\) on a probability space \(\p{\Omega, \mathcal A, \pmeasure}\). Letting \(\Sigma\p{V}\) denote the \(\sigma\)-algebra generated by a random variable \(V\), we assume throughout that \(\Sigma\p{\nvar{z}}\subset\Sigma\p{\nvar{y}}\subset\Sigma\p{\nvar{x}}\), such that \(\nvar{x}\) has a non-trivial dependence on \(\nvar{y}\), which in turn depends on \(\nvar{z}\). Let ${\tilde{X}}$ be a $\Sigma(X)$-measurable random variable and consider a random variable $V$ taking values in $\rset^d$, with $\Sigma(V)\subset \Sigma({\tilde{X}})$. For any $v\in \rset^d$, we define the random variable \({{\tilde{X}}^{\{V=v\}}:V^{-1}(\{v\})\to \rset}\) as the restriction of ${\tilde{X}}$ to the event \({\{V=v\}}\). When the joint distribution between ${\tilde{X}}$ and $V$ is clear from context, we shorten this notation to ${\tilde{X}}^{\{V=v\}} \defeq {\tilde{X}}^{\{v\}}$. In particular, for any \(\Sigma(\nvar{y})\)-measurable random variable \(\tilde Y:\Omega\to \rset^d\) it follows that \(\tilde X^{\{\tilde Y\}}:\Omega \to \rset\) is a \(\Sigma(X)\)-measurable  random variable.
By definition, $\nvarcond{x}{y}\equiv \nvar{x}$. 
}

\subsection{Assumptions}\label{sec:cva_setup_assumptions} 
In the context of financial risk estimation, the variable \(\nvar{z}\) in \eqref{eqn:cva_general_nested_prob} typically represents random market fluctuations over a short time horizon. As such, it is often possible to obtain sufficiently accurate samples of \(\nvar{z}\) with negligible cost. To simplify the discussion, we therefore assume that \(\nvar{z}\) can be sampled exactly, while emphasising that the methods which follow can be extended naturally to include approximation of \(\nvar{z}\). Given samples of \(\nvar{z}\), we assume that \(\nvar{x}\) and \(\nvar{y}\) can be approximated to sufficient accuracy, but with increasing cost as the error converges to zero. \final{For integers \(\ell,\, k\ge 0\) controlling refinement parameters, we denote approximations \(\nvar[k]{x}\approx \nvar{x}\) and \(\nvar[\ell]{y}\approx \nvar{y}\). We impose the following conditions on (approximations of) \(\nvar{x}\) and \(\nvar{y}\):}
\begin{assumption}\label{assumpt:cva_general_rv_convergence}
	For all \(\ell, k\ge 0\), \(y, z\in \rset^d\) and \(q\ge 2\), there exists a constant \(c > 0\), independent of \(\ell\) and \(k\) such that
	
	\begin{minipage}{0.48\linewidth}
		\[
			\begin{aligned}
				\textnormal{Cost}\p[\big]{\nvarcond[k]{x}{sy}} &\le c2^{k},\\
				\E[\Big]{\abs[\big]{\nvarcond[k][\ell]{x}{y} - \nvarcond[none][\ell]{x}{y}}^q} &\le c2^{-qk},\\
				\abs[\Big]{\E[\big]{\nvarcond[k][\ell]{x}{y} - \nvar{x}}} &\le c2^{-\min\br{\ell,k}},
			\end{aligned}
		\]
	\end{minipage}
	\begin{minipage}{0.47\linewidth}
		\[
		\begin{aligned}
		\textnormal{Cost}\p[\big]{\nvarcond[\ell]{y}{sz}} &\le c2^{\ell},\\
\st{\E[\Big]{\abs[\big]{\nvarcond[k][\ell]{x}{y} - \nvar[k]{x}}^q}} &\le \st{c2^{-q \ell}},\\
		\norm[\Big]{\E[\big]{\nvar[\ell]{y} - \nvar{y}}} &\le c2^{-\ell}.
		\end{aligned}
		\]
	\end{minipage}
\end{assumption}

Consider the variables \(\nvar{x}\) and \(\nvar{y}\) in \eqref{eqn:cva_vars}, depending on the solution of an SDE \eqref{eqn:cva_sde}.  \final{Let $\nvarcond[k][\ell]{x}{y}$ be sampled by discretising \eqref{eqn:cva_sde} using the Milstein scheme with step-size $h_\ell\propto 2^{-\ell}$ on $t\in [0,\tau]$ and $h_k\propto 2^{-k}$ on $t\in[\tau,T]$.} \MakeUppercase assumption\nobreakspace \ref {assumpt:cva_general_rv_convergence} holds, provided that \(\pi_{\tau, T}\) is a Lipschitz functional of the assets \nt{\(S\)}, whenever the coefficients \(\sdedrift\) and \(\sdevol\) of the SDE are Lipschitz continuous, twice differentiable and, combined with their first and (mixed) second order derivatives, satisfy linear growth conditions as specified \final{by Kloeden and Platen} in \cite[Theorem 10.3.5]{Kloeden:1999}.

The order of convergence specified by \MakeUppercase assumption\nobreakspace \ref {assumpt:cva_general_rv_convergence} enables the use of randomised MLMC techniques as \final{studied by Rhee and Glynn \cite{Rhee2015_Unbiased_estimation, rg12}, McLeish \cite{mcleish11}, Vihola \cite{mv18} and by Giles and Goda \cite{gg19}. Randomised MLMC estimators are used }in Section\nobreakspace \ref {sec:cva_unbiased_mlmc} to sample unbiased estimators of \({\nestE{1}}\) and \({\nestE{0}}\) with finite cost and variance, see Proposition\nobreakspace \ref {prop:cva_umlmc_moments}. 

We additionally impose the following condition on the function \(f\) within \eqref{eqn:cva_general_nested_prob}.
\begin{assumption}
	\label{assumpt:rec_f_decomposition}
	The function \(f:\rset\to\rset\) is \ntg{\(L_f\)-Lipschitz on \(\rset\)}, and twice differentiable on \(\rset\backslash \Gamma\) for some \(\Gamma\subset\rset\). The second derivative of \(f\) is uniformly bounded by \(L_f^{\prime} > 0\). That is,
	\begin{equation} \label{eqn:rec/f_der_bound}
	\begin{aligned}
	\sup_{\substack{u\in\rset\backslash\Gamma}}\abs[\big]{f^{\prime\prime}\p{u}} &\le L_f^\prime.
	\end{aligned}
	\end{equation}
	Moreover, there is \(\bar\delta,\bar\rho > 0\) such that \(0<x\le\bar\delta\) implies 
	\begin{equation}\label{eqn:rec/f_prob_bound}
	\prob[\Big]{\inf_{u\in\Gamma}\abs[\big]{{\nestE{1}} - u} \le x}\le \bar\rho x,
	\end{equation}
	where \(\bar\delta\) and \(\bar \rho\) are independent of \(x\).
\end{assumption}
Similar conditions to \MakeUppercase assumption\nobreakspace \ref {assumpt:rec_f_decomposition} are imposed \final{by Bourgey et al.. and by Giles and Szpruch in} \cite{Bourgey20,giles14antmilstein} to approximate averages of Lipschitz functionals, \(f\), of an underlying quantity of interest using antithetic MLMC differences based on a Taylor expansion of \(f\). This methodology is used in  Section\nobreakspace \ref {sec:cva_unbiased_mlmc} to approximate \({\nestE{0}}\). The global bound on the second derivative \eqref{eqn:rec/f_der_bound} can be used to bound a second-order Taylor expansion of \(f\) on regions where the function is smooth. The additional condition \eqref{eqn:rec/f_prob_bound} allows \(f\) to be non-\nt{continuously} differentiable on a set \(\Gamma\), \nt{which is essential when considering the CVA in \eqref{eqn:cva_master}}, provided we can linearly bound the probability that \({\nestE{1}}\) lies a distance \(x\) from \(\Gamma\).

\subsection{\nt{Nested} Monte Carlo Estimate}\label{sec:cva_hierarchical_mc}
\nt{To motivate our work and as a baseline for comparison,} we begin by considering an estimator for the quantity \(\pll\) defined by \eqref{eqn:cva_general_nested_prob} using a hierarchy of single-level Monte Carlo averages. In particular, due to the recursively nested structure of the problem and compounding approximation costs through several nested Monte Carlo estimates, we will see that accurate computation of \(\pll\) is expensive. We construct the estimator iteratively, starting from the innermost conditional expectation and working outward.

\final{For a fixed \(y\in \rset^d\), a Monte Carlo average for the innermost conditional expectation \({\nestE{1}}\), given \(\nvar{y} = y\), at level \(\kmc = \p{k_0, k_1}\in\nset^2\) is
\[
	\appE{1}{\kmc}\p{y} \defeq \frac{1}{N_{k_0}}\sum_{n_1=1}^{N_{k_0}}\nvarcond[k_1][none][n_1]{x}{sy},
\]
where \nt{we set} \(N_{k_0} = N_{0}2^{k_0}\) and \(\br{\nvarcond[k_1][none][n_1]{x}{sy}}_{n_1=1}^{N_{1,k_0}}\) are independent samples of \(\nvarcond[k_1][none]{x}{sy}\).} At approximation level \(k_1\), the cost of sampling \(\nvarcond[k_1]{x}{sy}\) is of order \(2^{k_1}\) under \MakeUppercase assumption\nobreakspace \ref {assumpt:cva_general_rv_convergence}. Hence, the sampling cost of \(\appE{1}{\kmc}\p{y}\) is of order \(N_{k_0}\textnormal{Cost}\p{\nvarcond[\ell][none]{x}{sy}} \propto 2^{k_0+k_1}\).

\final{Similarly, for $z\in \rset^d$, a (biased) Monte Carlo average to approximate \({\nestE{0}}\), given \(\nvar{z}=z\), at level \(\ellmc=\p{\ell_0,\ell_1}\in\nset^2\) is then given by
\[
	\appE{0}{\ellmc,\kmc}\p{\nvar{sz}} \defeq \frac{1}{N_{\ell_0}}\sum_{n_0=1}^{N_{\ell_0}} f\p[\Big]{\appE{1}{\kmc}^{\p{n_0}}\p[\big]{\nvarcond[\ell_1][none][n_0]{y}{sz}}},
\]
where \(\br{\nvarcond[\ell_1][none][n_0]{y}{sz},\, \appE{1}{\kmc}^{\p{n_0}}\p{\nvarcond[\ell_1][none][n_0]{y}{sz}}}_{n_0=1}^{N_{\ell_0}}\)  are independent samples of  \((\nvarcond[\ell_1][none]{y}{sz}, \appE{1}{\kmc}\p{\nvarcond[\ell_1]{y}{sz}})\), respectively. It follows under \MakeUppercase assumption\nobreakspace \ref {assumpt:cva_general_rv_convergence} that
\begin{equation}\label{eqn:nestEcost_mc}
\begin{aligned}
\textnormal{Cost}\p{\appE{0}{\ellmc,\kmc}\p{\nvar{sz}}} &= \Order[\bigg]{N_{\ell_0}\sup_{\nvar{sy}\in\rset^d} \p[\Big]{\textnormal{Cost}\p[\big]{\appE{1}{\kmc}\p{\nvar{sy}}} + \textnormal{Cost}\p[\big]{\nvarcond[\ell_1][none]{y}{sz}}}}\\
&= \Order{2^{\ell_0 + k_0 + k_1} + 2^{\ell_0+\ell_1}}.
\end{aligned}
\end{equation}
}

Using the above nested Monte Carlo estimator for \({\nestE{0}}\), a \nt{nested} Monte Carlo estimate for \(\prob{{\nestE{0}} > \lloss}\) is given by
\[
	\pll_{M, \ellmc, \kmc} = \frac{1}{M}\sum_{m=1}^{M}\I{\appE{0}{\ellmc,\kmc}^{\p{m}}\p{\nvar[none][m]{z}} > \lloss},
\]
where \(\br{\p{\nvar[none][m]{z},\appE{0}{\ellmc,\kmc}^{\p{m}}\p{\nvar[none][m]{z}}}}_{m=1}^M\) are independent realisations of \(\p{\nvar{z},\appE{0}{\ellmc,\kmc}\p{\nvar{z}}}\), respectively. Following \eqref{eqn:nestEcost_mc}, the sampling cost of \st{\(\pll_{M, \ellmc, \kmc}\)} is  given by 
\[
\st{\textnormal{Cost}\p{\pll_{M, \ellmc, \kmc}} = 
\Order[\big]{M\p{2^{\ell_0 + k_0 + k_1} + 2^{\ell_0+\ell_1}}}.}
\]
Meanwhile, the mean square error of the approximation \(\pll_{M, \ellmc, \kmc}\) can be decomposed into a bias and statistical error through \st{the decomposition}
\[
	\st{\E[\big]{\p{\pll - \pll_{M, \ellmc, \kmc}}^2} = \abs[\big]{\E{\pll - \pll_{M, \ellmc, \kmc}}}^2 + \var{\pll_{M, \ellmc, \kmc}}.}
\]
By following a similar argument to \final{Bujok et al..} \cite[Proposition 1]{BujokK2015}, under \MakeUppercase assumption\nobreakspace \ref {assumpt:cva_general_rv_convergence} and assuming that the joint distribution of \({\nestE{0}}\) and \(\appE{0}{\ellmc,\kmc}\p{\nvar{z}} \) admits a density \(\rho\p{u,\hat u}\) which is twice differentiable in \(u\) for all \(\ellmc, \kmc\), one can show that \(\abs{\E{\pll - \pll_{M, \ellmc, \kmc}}}\) is of order \(2^{-\min\br{\ell_0, \ell_1, k_0,k_1}}\). The hierarchical estimator discussed in the following section does not require a bound on the estimation bias of this form and therefore \nt{we do not include a rigorous proof of this bound}. Furthermore, \nt{it can be shown} that \(\var{\pll_{M, \ellmc, \kmc}} = \Order{M^{-1}}\). Therefore, to ensure \(\E{\p{\pll - \pll_{M, \ellmc, \kmc}}^2} \le \tol^{2}\), we require \(M = \Order{\tol^{-2}}\), and  \ntg{\(2^{-\min\br{\ell_0, \ell_1, k_0,k_1}}=\Order{\tol}\)}. Hence, the cost of attaining a root mean square error \(\tol\) for the estimator \(\pll_{M, \ellmc, \kmc}\) is of order \(\tol^{-5}\).
This scaling of computational cost makes it extremely expensive to attain sufficiently accurate estimates of \(\pll\) using \(\pll_{M, \ellmc, \kmc}\), particularly for the root finding problem of computing \(\lloss\) in \eqref{eqn:cva_general_nested_prob}, which requires computing several estimates of \(\pll_{M, \ellmc, \kmc}\) with different loss thresholds \(\lloss\). This motivates the development of more advanced estimators of \(\pll\), with a computational cost which scales more efficiently with the designated error tolerance \(\tol\).

\subsection{\nt{Nested} Unbiased MLMC Estimate}\label{sec:cva_unbiased_mlmc}
To improve the computational cost, we first consider the use of recursive randomised MLMC estimation \ntg{\cite{Rhee2015_Unbiased_estimation,mcleish11,rg12}} to construct an estimator of \({\nestE{0}}\) with a cost which is lower than that in \eqref{eqn:nestEcost_mc}. We begin by considering the innermost expectation \({\nestE{1}}\). \final{Consider the multilevel differences defined by
\[
	\dl[k]\nvar{x} \defeq
	\begin{cases}
		\nvar[k]{x} - \nvar[k-1]{x} & k\ge 1\\
		\nvar[0]{x}	&	k=0,
	\end{cases}
\] 
so that under \MakeUppercase assumption\nobreakspace \ref {assumpt:cva_general_rv_convergence} we have
\(
	{\nestE{1}} = \E{\sum_{k=0}^\infty \dl[k]\nvar{x}\given \nvar{y}}.
\)}
Then, for \(\zeta_1 > 1\), define a random non-negative integer \(\rlev{1}\), which is independent of \(\br{\nvar[k]{x}}_{k=0}^\infty\), \nt{satisfying
\(
	p_{1,k} \defeq \prob{\rlev{1} = k}  =  \p{1 - 2^{-\zeta_1}}2^{\ntg{-}\zeta_1 k}.
\)}
It follows that \cite{Rhee2015_Unbiased_estimation,GilesHajiAli:2019sampling}
\begin{equation}\label{eqn:cva_ub_innermost}
	\begin{aligned}
		{\nestE{1}}&=\E[\bigg]{\sum_{k=0}^{\infty} \dl[k]\nvar{x}\given \nvar{y}}\\
		 &= \E[\bigg]{\sum_{k=0}^{\infty} \p[\big]{\dl[k]\nvar{x}p_{1, k}^{-1}} p_{1,k}\given \nvar{y}}\\
		  &= \E[\bigg]{\dl[\rlev{1}]\nvar{x} p_{1,\rlev{1}}^{-1}\given \nvar{y}},
	\end{aligned}
\end{equation}
\ntg{where \(p_{1,\rlev{1}}\) denotes the probability mass function \(p_{1,k}\) evaluated at the random variable \(\rlev{1}\).}
An unbiased Monte Carlo estimate of \({\nestE{1}}\), \final{given $\nvar{y}=\nvar[\ell]{sy}\in\rset^d$ is thus given by 
\[
	\appEub{1}{\ell}\p{\nvar[\ell]{sy}} = \frac{1}{N_{1,\ell}}\sum_{n=1}^{\nt{N_{1,\ell}}} \dl[\rlev{1}^{\p{n}}]\nvarcond[none][\ell][n]{x}{sy} p_{1,\rlev{1}^{\p{n}}}^{-1},
\]
where \(N_{1,\ell} \defeq N_{1,0}2^{\ell}\) and the random variables \(\br{\p{\rlev{1}^{\p{n}}, \dl[\rlev{1}^{\p{n}}]\nvarcond[none][\ell][n]{x}{sy}}}_{n=1}^{N_{1,\ell}}\) are independent samples of \(\p{\rlev{1}, \dl[\rlev{1}]\nvarcond[none][\ell]{x}{sy}}\).} Under the conditions in \MakeUppercase assumption\nobreakspace \ref {assumpt:cva_general_rv_convergence}, it follows that \(\appEub{1}{\ell}\p{\nvar[\ell]{sy}}\) has finite variance and an expected sampling cost of order \(N_{1,\ell}\) provided \(1<\zeta_1<2\) \cite{GilesHajiAli:2019sampling,Rhee2015_Unbiased_estimation}.

Following the approach in \cite{Bourgey20,GilesHajiAli:2018,BujokK2015}, we use the unbiased Monte Carlo estimates of \final{\({\nestE{1}}\)} to construct an antithetic multilevel difference for the term \final{\(f\p{{\nestE{1}}}\)} appearing within \final{the definition of \({\nestE{0}}\)}. Specifically, we define
\begin{equation}\label{eqn:dlantf}
	\dl[\ell]^{\p{\textnormal{UB}}} f = f\p[\Big]{\antEfub[\ell][1]\p{\nvar[\ell]{y}}} - \frac{1}{2}\sum_{i=0}^1f\p[\Big]{\antEcub[i][\ell-1][1]\p{\nvar[\ell-1]{y}}},
\end{equation}
where,\final{ given \(\nvar[\ell-1]{y}\), \(\br{\antEcub[i][\ell-1][1]\p{\nvar[\ell-1]{y}}}_{i=0}^1\) are conditionally independent realisations of the random variable} \(\appEub{1}{\ell-1}\p{\nvar[\ell-1]{y}}\). \final{Specifically, for $\nvar[\ell-1]{y} = \nvar[\ell-1]{sy}\in\rset^d$ we define}
\[
 \antEcub[i][\ell-1][1]\p{\nvar[\ell-1]{sy}} = \frac{1}{N_{\nt{1, }\ell-1}}\sum_{n=1}^{N_{\nt{1, }\ell-1}}\dl[\rlev{1}^{\p{n}}]\nvarcond[none][\ell-1][n+iN_{\ell-1}]{x}{sy} p_{1,\rlev{1}^{\p{n}}}^{-1},
\]
where the samples \nt{\(\dl[\rlev{1}^{\p{n}}]\nvarcond[none][\ell-1][n]{x}{sy}\) \ntg{should be} correlated to \(\dl[\rlev{1}^{\p{n}}]\nvarcond[none][\ell][n]{x}{sy}\) which are used to compute \(\antEfub[\ell][1]\p{\nvar[\ell]{sy}}\),} in the sense of \MakeUppercase assumption\nobreakspace \ref {assumpt:cva_general_rv_convergence}. \ntg{In particular,} for a given \(k\), the \(\leb^q\)-error between both terms can be bounded by
\[
\begin{aligned}
	&\E[\Big]{\abs[\big]{\dl[k]\nvarcond[none][\ell][n]{x}{y} - \dl[k]\nvarcond[none][\ell-1][n]{x}{y}}^q}\\
	&= \E[\Big]{\abs[\big]{\nvarcond[k][\ell][n]{x}{y} - \nvarcond[k-1][\ell][n]{x}{y} - \nvarcond[k][\ell-1][n]{x}{y} + \nvarcond[k-1][\ell-1][n]{x}{y}}^q}\\
	&\le a2^{-\ntg{q}\max\br{\ell, k}},
\end{aligned}
\]
\ntg{where we use Jensen's inequality to bound 
\begin{equation}
	\label{eqn:double_difference_lgk}
\begin{aligned}
&\E[\Big]{\abs[\big]{\dl[k]\nvarcond[none][\ell][n]{x}{y} - \dl[k]\nvarcond[none][\ell-1][n]{x}{y}}^q}\\
&\le 2^{q-1}\bigg(\E[\Big]{\abs[\big]{\nvarcond[k][\ell][n]{x}{y} - \nvarcond[k-1][\ell][n]{x}{y}}^q}
+\E[\Big]{\abs[\big]{\nvarcond[k][\ell-1][n]{x}{y} + \nvarcond[k-1][\ell-1][n]{x}{y}}^q}\bigg),
\end{aligned}
\end{equation}
when \(k \ge \ell\), and again to bound
\begin{equation}
	\label{eqn:double_difference_kgl}
\begin{aligned}
&\E[\Big]{\abs[\big]{\dl[k]\nvarcond[none][\ell][n]{x}{y} - \dl[k]\nvarcond[none][\ell-1][n]{x}{y}}^q}\\
&\le 2^{q-1}\bigg(
\E[\Big]{\abs[\big]{\nvarcond[k][\ell][n]{x}{y} - \nvarcond[k][\ell-1][n]{x}{y}}^q} +
\E[\Big]{\abs[\big]{\nvarcond[k-1][\ell][n]{x}{y} - \nvarcond[k-1][\ell-1][n]{x}{y}}^q}
\bigg),
\end{aligned}
\end{equation}
when \(\ell>k\). It follows from \MakeUppercase assumption\nobreakspace \ref {assumpt:cva_general_rv_convergence} that \eqref{eqn:double_difference_lgk} is bounded by \(c2^{-k}\) and \eqref{eqn:double_difference_kgl} by \(c2^{-\ell}\).}
In particular, on the event \(\nvar[\ell]{y} = \nvar[\ell-1]{y} = y\), it follows that
\[
	\antEfub[\ell][1]\p{y} - \frac{1}{2}\sum_{i=0}^1 {\antEcub[i][\ell-1][1]\p{y}} = 0.
\]
Therefore, on a linearisation of \(f\), it follows that \(\dl[\ell]^{\p{\textnormal{UB}}} f\) is dominated by the approximation error between \(\nvar[\ell]{y}\) and \(\nvar[\ell-1]{y}\). This fact can be used to prove Lemma\nobreakspace \ref {lem:cva_general_antithetic_result} \nt{in the appendix}, which can be used to establish \nt{the} convergence of the moments of \(\dl[\ell]^{\p{\textnormal{UB}}} f\).

Randomised MLMC techniques can be applied recursively to create an unbiased Monte Carlo estimator of \({\nestE{0}}\), based on a randomisation of the \nt{approximation} level \(\ell\) used for the antithetic differences \(\dl[\ell]^{\p{\textnormal{UB}}} f\). Let \(\rlev{0}\) be a random, non-negative, integer  with mass function \nt{ 
\(
	p_{0,\ell}\defeq \prob{\rlev{0} = \ell} = \p{1-2^{-\zeta_0}}2^{-\zeta_0\ell}.
\)}
Then, for $\nvar{sz}\in\rset^d$, consider the random variable
\begin{equation}\label{eqn:delta_unbiased}
	\dpz^{\{\nvar{sz}\}} \defeq  \dl[\rlev{0}]^{\p{\textnormal{UB}}} f\p[\big]{\nvarcond[\kappa_0]{y}{sz}, \nvarcond[{\kappa_0-1}]{y}{sz}} p_{0,\rlev{0}}^{-1},
\end{equation}
\ntg{where \(p_{0,\rlev{0}}\) denotes the probability mass \(p_{0,k}\) evaluated at the random variable \(\rlev{0}\).} \final{In what follows, we use the notation $\dpz = \dpz^{\{Z\}}$ to refer to the random variable in \eqref{eqn:delta_unbiased} without restricting to the event $\nvar{z}=\nvar{sz}$.}  \ntg{By a similar calculation to \eqref{eqn:cva_ub_innermost},  \(\dpz\)} satisfies
\(
	{\nestE{0}} = \E{\dpz\given \nvar{z}}.
\) 
In particular, consider the unbiased Monte Carlo estimate of \({\nestE{0}}\), given $\nvar{z}=\nvar{sz}$, \final{defined by
\begin{equation}\label{eqn:cva_umlmc_middle}
\appEub{0}{\ell}(\nvar{sz}) = \frac{1}{N_{0,\ell}}\sum_{n=1}^{N_{0,\ell}} \dpz^{\{z\}, (n)},
\end{equation}
where $\{\dpz^{\{z\}, (n)}\}_{n=1}^{N_{0,\ell}}$ are independent samples of $\dpz^{\{z\}}$ for \(N_{0,\ell}\defeq N_{0,0}2^{\gamma\ell}\), \nt{with} \(\gamma>0\) \nt{fixed}.} 
The following result, which is proved in Appendix\nobreakspace \ref {app:proofs}, can be used to bound the sampling cost and variance of a sample from \eqref{eqn:cva_umlmc_middle}, with fixed values of \(\zeta_0\) and \(\zeta_1\). 
\begin{proposition}\label{prop:cva_umlmc_moments}
	Let \MakeUppercase assumptions\nobreakspace \ref {assumpt:cva_general_rv_convergence} and\nobreakspace  \ref {assumpt:rec_f_decomposition} hold and
	consider the variable \(\dpz\) defined by \eqref{eqn:delta_unbiased}, where we fix \(\zeta_0 = 17/16\) and \(\zeta_1 = 3/2\). Then, \(\dpz\) has finite expected sampling cost and variance. Moreover, \(\E{\abs{\dpz}^{q_0}}<\infty\) for all
	\(
	q_0 < 23/11.
	\)
\end{proposition}
\nt{As a consequence of Proposition\nobreakspace \ref {prop:cva_umlmc_moments}, for the same values of \(\zeta_0, \zeta_1\) and \(q_0\), it follows that the Monte Carlo estimate \eqref{eqn:cva_umlmc_middle} has a sampling cost of order \(N_{0,\ell}\) and finite \(q_0^\text{th}\) moment.}
\begin{remark}
	By the linearity of expectation, the approach in this section can be applied naturally in the case where \final{\({\nestE{0}}\) or \({\nestE{1}}\)} are replaced with a linear combination of expectations of the same form, each satisfying \MakeUppercase assumptions\nobreakspace \ref {assumpt:cva_general_rv_convergence} and\nobreakspace  \ref {assumpt:rec_f_decomposition}. This fact \nt{will be} useful when considering the CVA problem in Section\nobreakspace \ref {sec:cva_cva_capital}.
\end{remark}

\nt{More generally, \MakeUppercase assumption\nobreakspace \ref {assumpt:cva_general_rv_convergence} could be generalised to include  Euler-Maruyama type convergence rates by relaxing the convergence rates \(2^{-q\ell}\) and \(2^{-qk}\) to \(2^{-q\ell/2}\) and \(2^{-qk/2}\) therein, however the conclusion of Proposition\nobreakspace \ref {prop:cva_umlmc_moments} will no longer hold true. In this setting, it is possible to construct a hierarchy of nested biased MLMC estimates for \(\final{{\nestE{0}}}\) and \(\final{{\nestE{1}}}\) \cite{spence23}.}

\subsection{Adaptive MLMC for Probabilities}\label{sec:adapt_mlmc}
Using the estimator \eqref{eqn:delta_unbiased}, the problem \eqref{eqn:cva_general_nested_prob} can be reduced to
\begin{equation*}\label{eqn:cva_simplified_nested_prob}
	\begin{aligned} 
		\pll &= \prob[\big]{{\nestE{0}} > \lloss}= \E[\big]{\I{{\nestE{0}} > \lloss}},\\
		{\nestE{0}} &= \E[\big]{\nvarcond{d}{z}\given \nvar{z}},
	\end{aligned} 
\end{equation*}
where $\Sigma(Z)\subset \Sigma(V)$, and \(\nvarcond{d}{z}\) can be sampled with finite expected sampling cost. At level \(\ell\ge 0\), we consider the Monte Carlo estimate 
\(
	\appEub{0}{\ell}\p{\nvar{sz}}
\)
defined in \eqref{eqn:cva_umlmc_middle}. Since \MakeUppercase assumption\nobreakspace \ref {assumpt:rec_f_decomposition} does not hold for the discontinuous function \(\I{x>\lloss}\), \nt{the methods and analysis of the previous section are not applicable to construct an unbiased estimator for \(\pll\)} which has both finite sampling cost and variance. Instead, we consider a truncated (biased) MLMC estimate \cite{Giles2008MLMC} defined using the multilevel differences \cite{GilesHajiAli:2018}
\begin{equation}\label{eqn:dhl}
	\dhl \defeq 
	\begin{cases}
		\I{\appEub{0}{0}\p{\nvar[none]{z}} > \lloss} & \ell = 0\\
		\I{\appEub{0}{\ell}\p{\nvar[none]{z}}>\lloss} - \I{\appEub{0}{\ell-1}\p{\nvar[none]{z}}>\lloss} & \ell>0,
	\end{cases}
\end{equation}
through
\begin{equation}\label{eqn:cva_pll_mlmc}
\begin{aligned}
\pll\approx \pll^{\p{\textnormal{MLMC}}}_{\Mvec, L} \defeq\sum_{\ell=0}^L \frac{1}{M_\ell}\sum_{m=1}^{M_\ell}\dhl^{\p{m}},
\end{aligned}
\end{equation}
for a truncation level \(L\in\nset\) and inner sample sizes \(\Mvec=\p{M_0,\dots, M_L}\in \nset^{L+1}\) and where \(\br{\dhl^{\p{m}}}_{m=1}^{M_\ell}\) are independent samples of \(\dhl\) for \(0\le \ell\le L\).  By the linearity of expectation, \(\pll^{\p{\textnormal{MLMC}}}_{\Mvec, L}\) is an unbiased estimate of \(\prob{\appEub{0}{L}\p{\nvar{z}} > \lloss}\approx \pll\). Furthermore, for large \(\ell\), the variance of \(\dhl\) becomes small, requiring fewer samples at costly levels \(\ell\gg 1\) to control the variance of the estimator \(\pll^{\p{\textnormal{MLMC}}}_{\Mvec, L}\). From standard MLMC complexity analysis of \final{Cliffe et al.. \cite{Cliffe:2011} and Giles \cite{giles:2015review}}, since the cost of sampling \(\dhl\) is of order \(N_{0,\ell}\) by Proposition\nobreakspace \ref {prop:cva_umlmc_moments}, for a specified root mean square error of \(\tol\) it is optimal to allocate \(\Mvec = \Mvec^\star\), defined component-wise by \(M_\ell^\star \propto \tol^{-2}\sqrt{V_\ell N_{0,\ell}^{-1}}\sum_{k=0}^L \sqrt{V_k N_{0,k}}\), where \(V_\ell\defeq \var{\dhl}\). For this choice of sample sizes, the cost of the MLMC estimate is
\begin{equation}\label{eqn:mlmc_cost}
	\text{Cost}\p{\pll^{\p{\textnormal{MLMC}}}_{\Mvec^\star, L};\tol} \propto \tol^{-2}\p[\bigg]{\sum_{\ell=0}^L \sqrt{V_\ell N_{0,\ell}}}^2,
\end{equation}
where the truncation level \(L\) should be chosen such that the bias \st{is bounded by} \({\abs{\E{\pll-\pll^{\p{\textnormal{MLMC}}}_{\Mvec, L}}} \le \tol/2}\). From \cite[Proposition 1]{Gordy:2010}, under the assumption that the joint distribution of \({\nestE{0}}\) and \(\appEub{0}{\ell}\p{\nvar{z}}\) exists and is twice differentiable in \({\nestE{0}}\), it follows that \(\abs{\E{\dhl}} = \Order{N_{0,\ell}^{-1}}\). In particular, to meet \nt{a} specified \nt{root mean square error tolerance, \(\tol\),} one should scale \(L\) \ntg{such that \(2^{-L} = \Order{\tol}\)}. In practice, both the bias \(\abs{\E{\dhl}}\) and variances \(V_\ell\) must be approximated from the available samples of \(\dhl\) in order to estimate the optimal truncation level \(L\) and \nt{\(\Mvec^\star\)}. 

To study the asymptotic behaviour of this estimator, we impose an additional condition on the approximations of \({\nestE{0}}\) \cite[Assumptions 1.3 and 1.4]{hajiali2021adaptive}. Within this assumption and in what follows, we introduce an abstract sequence of positive-valued random variables \(\br{\sigma_\ell}_{\ell\in\nset}\), which can be used to improve the concentration properties of the normalized statistic \(\sigma_\ell^{-1}\appEub{0}{\ell}\p{\nvar{z}} \). 

\begin{assumption}\label{assumpt:admlmc}
	For some \(p>2\) and \nt{a} sequence \(\br{\sigma_{\ell}}_{\ell\in\nset}\) of positive-valued random variables, the \(p^\text{th}\) centralized moment of \(\appEub{0}{\ell}\p{\nvar{z}}\), normalized by \(\sigma_\ell N_{0,\ell}^{-1/2}\), is finite. That is,
	\begin{equation}\label{eqn:assumpt_ad_moment}
		\lim_{\ell_0\to\infty}\sup_{\ell>\ell_0}\E[\bigg]{\abs[\Big]{\frac{\appEub{0}{\ell}\p{\nvar{z}} - {\nestE{0}}}{\sigma_\ell N_{0,\ell}^{-1/2}}}^p} < \infty.
	\end{equation}
	Moreover, there exists \(\delta_0, \rho_0 > 0\) such that
	\begin{equation}\label{eqn:assumpt_ad_cdf}
		\sup_{\substack{\ell\ge0\\ 0<x\le \delta_0}} \prob[\bigg]{\abs[\Big]{\frac{\appEub{0}{\ell}\p{\nvar{z}}-\lloss}{\sigma_\ell}} < x} \le \rho_0 x.
	\end{equation}
\end{assumption}

Within \MakeUppercase assumption\nobreakspace \ref {assumpt:admlmc}, the condition \eqref{eqn:assumpt_ad_moment} controls the \(\leb^p\)-error, normalized by \(\sigma_\ell\), made in approximating the nested expectation \({\nestE{0}}\) by the unbiased MLMC estimate \eqref{eqn:cva_umlmc_middle}. 
Under the conditions and result of Proposition\nobreakspace \ref {prop:cva_umlmc_moments}, it follows that this condition holds for any constant \(\sigma_\ell\equiv \sigma\in \rset\) and any \(p < 23/11\). Alternatively, one can consider setting \(\sigma_\ell^2\) to be the sample variance of the samples used to compute \(\appEub{0}{\ell}\p{\nvar{z}}\),
\begin{equation}\label{eqn:sigma_sd}
\sigma_\ell^2 \defeq \frac{1}{N_{0,\ell}}\sum_{n=1}^{N_{0,\ell}} \p[\big]{\nvarcond[none][none][n]{d}{z} - \appEub{0}{\ell}\p{\nvar{z}}}^2.
\end{equation}
In this context, the random variable \(\p{\appEub{0}{\ell}\p{\nvar{z}} - {\nestE{0}}}{\sigma_\ell^{-1} N_{0,\ell}^{1/2}}\) is equivalent to Student's test statistic for the samples \(\br{\nvarcond[none][none][n]{d}{z}}_{n=1}^{N_{0,\ell}}\).  The samples \(\nvarcond{d}{z}\) satisfy the hypothesis of the central limit theorem from Proposition\nobreakspace \ref {prop:cva_umlmc_moments} and, consequently, one can leverage results from the theory of self-normalising sums, \final{as discussed in Pe{\~n}a et al..} \cite{pena2008self} to analyse the condition \eqref{eqn:assumpt_ad_moment}. In particular,  under the assumption that there exists \(\rho_0, \ \lambda > 0\)  such that
\begin{equation}\label{eqn:snn_cond}
{\sup_{\substack{\delta\in\rset, z\in\rset^d}}}\prob[\bigg]{\abs[\big]{\nvarcond{d}{z} - {\nestE{0}} - \delta} \le \abs{\delta}x\given \nvar{z} = z} < \rho_0 x^\lambda,
\end{equation}
it follows from \final{Jonsson} \cite[Theorem 6.2]{JonssonFredrik2011} that \eqref{eqn:assumpt_ad_moment} holds for any \(p<\infty\). 
In addition, the bound \eqref{eqn:assumpt_ad_cdf} controls the probability of sampling an approximation for \({\nestE{0}}\) which lies close to the discontinuity at \(\lloss\).  
Under \MakeUppercase assumption\nobreakspace \ref {assumpt:admlmc}, it follows from \cite[Proposition 2.3, Corollary 2.10]{hajiali2021adaptive} that \(V_\ell\) is of order \(N_{0,\ell}^{-p/2\p{p+1}}\) and therefore \nt{\(\text{Cost}\p{\pll^{\p{\textnormal{MLMC}}}_{\Mvec^\star, L};\tol}\), for \ntg{\(N_{0,L}^{-1} = \Order{\tol}\)}}, is of order \(\tol^{-3+p/2\p{p+1}}\). While the hierarchical approach improves upon the order \(\tol^{-5}\) cost discussed in Section\nobreakspace \ref {sec:cva_hierarchical_mc}, the cost still scales worse than order \(\tol^{-2}\), which would be observed using standard Monte Carlo simulation if one could sample \({\nestE{0}}\) exactly with finite cost. The increased cost for the estimator \(\pll^{\p{\textnormal{MLMC}}}_{\Mvec, L}\) can be attributed to the fact that an order 1 error in \(\dhl\) is made whenever a pair of samples \(\appEub{0}{\ell}\p{\nvar[none][\ell,m]{z}}\), \(\appEub{0}{\ell-1}\p{\nvar[none][\ell, m]{z}}\) lie on opposite sides of \(\lloss\), inflating the variances \(V_\ell\) \cite{GilesHajiAli:2018,hajiali2021adaptive}. \final{Solutions to this problem are well studied within the MLMC literature and include numerical smoothing as discussed by Bayer et al.. \cite{bayerchiheb:2020,bcr:23}, path-branching as discussed by Giles and Haji-Ali \cite{gh22b} and nested Quasi-Monte Carlo sampling as discussed by Xu et al.. \cite{xu2020}. A comprehensive review of MLMC for nested simulation can be found in the review paper by Giles  \cite{g23discont}.}

In the present work, we consider adaptive \final{multilevel sampling as discussed in  \cite{GilesHajiAli:2018,GilesHajiAli:2019sampling, Brodie:2011}} by refining approximations of \({\nestE{0}}\) which are relatively close to the discontinuity \(\lloss\), thereby reducing the chance that \(\dhl\) is non-zero. Approximations of \({\nestE{0}}\) which lie relatively far from \(\lloss\) do not require significant refinement, allowing us to control the computational cost. \final{Similar techniques for computing failure probabilities have been studied by Elfverson et al.. in \cite{Elfverson:2016selectiverefinement}, and Dodwell et al.. \cite{Dodwell:2021refinement}}. Specifically, we introduce a random refinement parameter \(\eta_\ell\), taking values in \(\br{0,\dots, \ell}\), to the discretisation \(\appEub{0}{\ell+\eta_\ell}\p{\nvar{z}}\). For \(k\in\br{0,\dots, \ell}\) and \(1 < \refp < 2\), the value of \(\eta_\ell\) is chosen according to the relation \cite{GilesHajiAli:2018,hajiali2021adaptive}
\begin{equation}\label{eqn:choice_k}
	\eta_\ell = k \iff 
	\begin{cases}
		\abs[\bigg]{\frac{\appEub{0}{\ell+m}\p{\nvar{z}}-\lloss}{\sigma_{\ell+m}}} \le 	N_{0,\ell}^{\p{1-\refp}/\refp}N_{0,m}^{-1/\refp} & \text{for all } 0\le m \le k-1\\
		\abs[\bigg]{\frac{\appEub{0}{\ell+k}\p{\nvar{z}}-\lloss}{\sigma_{\ell+k}}} > 	N_{0,\ell}^{\p{1-\refp}/\refp}N_{0,k}^{-1/\refp} & \text{if } k < \ell,
	\end{cases}
\end{equation}
where \(\br{\sigma_\ell}_{\ell\in\nset}\) is as in \MakeUppercase assumption\nobreakspace \ref {assumpt:admlmc}.
The value of \(\eta_\ell\) increases according to uncertainty in the sign of \({\nestE{0}} - \lloss\). The parameter \(\refp\in\p{1,2}\) controls the strictness of the refinement procedure. Larger values of \(\refp\) lead to less strict refinements, with less cost but greater chance of error. When \(\sigma_{\ell}\) is chosen to be the sample standard deviation as in \eqref{eqn:sigma_sd}, the refinements outlined in \eqref{eqn:choice_k} can be thought of as conducting a one-sample t-test for the null hypothesis \({\nestE{0}} = \lloss\). The procedure for generating adaptive Monte Carlo approximations for \({\nestE{0}}\) is outlined in Algorithm\nobreakspace \ref {alg:adaptivity}, which is based on \cite[Algorithm 1]{hajiali2021adaptive}. An adaptive MLMC estimate of \(\pll\) is constructed as in \(\eqref{eqn:cva_pll_mlmc}\) using the multilevel differences
\begin{equation}\label{eqn:dhl_adapt}
	\dhl \defeq 
	\I{\appEub{0}{\ell+\eta_\ell}\p{\nvar[none]{z}}>\lloss} - \I{\appEub{0}{\ell-1 + \eta_{\ell-1}}\p{\nvar[none]{z}}>\lloss},
\end{equation}
when \(\ell>0\). The key convergence properties of the adaptive hierarchical estimator is summarised in the following result, which is proved in Appendix\nobreakspace \ref {app:proofs}.
\begin{theorem}
	\label{thm:ad_mlmc_hierarchy} 
	Let \MakeUppercase assumption\nobreakspace \ref {assumpt:admlmc} hold for some \(p\ge 2\) and random variables \(\sigma_\ell\) which can be sampled with order \(2^{\ell}\) cost. The MLMC estimator \(\pll^{\p{\textnormal{MLMC}}}_{\Mvec, L}\) defined through \eqref{eqn:cva_pll_mlmc} and \eqref{eqn:dhl_adapt}, with adaptive refinements obtained through Algorithm\nobreakspace \ref {alg:adaptivity} with \(1<\refp<2\) satisfies 
	\[
		\var{\dhl} = \Order{N_{0,\ell}^{-p/\p{p+2}}} \quad \text{and} \quad
		 \E{N_{0, \ell+\eta_\ell}} = \Order{N_{0,\ell}}.
	\] 
	Consequently,  the cost of sampling \(\pll^{\p{\textnormal{MLMC}}}_{\Mvec^\star, L}\) within a specified root mean square error \(\tol\) using optimal sample sizes defined by \(\Mvec^\star\) can be characterized as follows:
	\begin{itemize}
		\item \nt{Under the additional assumption that} \(\abs{\E{\dhl}} = \Order{N_{0,\ell}^{-1}}\), the cost of obtaining a root mean square error \(\tol\) satisfies \(\text{Cost}\p{\pll^{\p{\textnormal{MLMC}}}_{\Mvec^\star, L};\tol} = \Order{\tol^{-2\p{p+3}/\p{p+2}}}\) .
		\item Otherwise, the cost can be bounded by \(\text{Cost}\p{\pll^{\p{\textnormal{MLMC}}}_{\Mvec^\star, L};\tol} = \Order{\tol^{-2\p{p+2}/p}}\). 
	\end{itemize}
\end{theorem}

\begin{remark}\label{rem:thm_p_inf}
	Theorem\nobreakspace \ref {thm:ad_mlmc_hierarchy} implies the asymptotic cost of computing \(\pll^{\p{\textnormal{MLMC}}}_{\Mvec^\star, L}\) to an error tolerance \(\tol\) improves as \(p\to\infty\) within \MakeUppercase assumption\nobreakspace \ref {assumpt:admlmc}. Proposition\nobreakspace \ref {prop:cva_umlmc_moments} shows that the conditions hold with constants \(\sigma_\ell\equiv \sigma\) for any \(p<23/11\). Alternatively, choosing \(\sigma_\ell\) to be the sample standard deviation \eqref{eqn:sigma_sd} and under the condition \eqref{eqn:snn_cond}, the hypothesis of Theorem\nobreakspace \ref {thm:ad_mlmc_hierarchy} holds for any \(p<\infty\).
	 Taking \nt{\(p\) to be arbitrarily large}, one obtains  \(\textnormal{Cost}\p{\pll^{\p{\textnormal{MLMC}}}_{\Mvec^\star, L};\tol} = \Order{\tol^{-2-\delta}}\) for any \(\delta>0\).
\end{remark}

\begin{center}
	\begin{algorithm}
		\KwIn{Discretization level \(\ell\), Sample \(\nvar{z}\), Threshold \(\lloss\), Refinement parameter \(\refp\in\p{1,2}\), Sequence \(\br{\sigma_\ell}_{\ell\in\nset}\) satisfying \MakeUppercase assumption\nobreakspace \ref {assumpt:admlmc}}
		\KwResult{Adaptively refined approximation \(\appEub{0}{\ell+\eta_\ell}\p{\nvar{z}}\)}
		Set \(\eta_\ell = 0\)\;
		Generate \(N_{0,\ell}\) independent samples of \(\br{\nvarcond[none][none][n]{d}{z}}_{n=1}^{N_{0,\ell}}\) given \(\nvar{z}\) through the construction \eqref{eqn:delta_unbiased}\;
		Set \(\appEub{0}{\ell}\p{\nvar{z}} = N_{0,\ell}^{-1}\sum_{n=1}^{N_{0,\ell}}\nvarcond[none][none][n]{d}{z}\)\;
		Sample \(\sigma_\ell\) given \(\br{\nvarcond[none][none][n]{d}{z}}_{n=1}^{N_{0,\ell}}\)\;
		\While{
			\(\abs{\appEub{0}{\ell+\eta_\ell}\p{\nvar{z}}-\lloss} \le \sigma_{\ell+\eta_\ell}N_{0,\ell}^{\p{1-\refp}/\refp}N_{0,\eta_\ell}^{-1/\refp}\) and \(\eta_\ell < \ell\)
		}{
			Generate  new, independent samples \(\br{\nvarcond[none][none][n]{d}{z}}_{n=N_{0,\ell+\eta_\ell}}^{N_{0,\ell+\eta_\ell+1}}\) given \(\nvar{z}\)\;
			Set \(\appEub{0}{\ell+\eta_\ell+1}\p{\nvar{z}} = N_{0,\ell+\eta_\ell+1}^{-1}\sum_{n=1}^{N_{0,\ell+\eta_\ell+1}}\nvarcond[none][none][n]{d}{z}\)\;
			Sample \(\sigma_{\ell+\eta_\ell}\) given \(\br{\nvarcond[none][none][n]{d}{z}}_{n=1}^{N_{0,\ell+\eta_\ell}}\)\;
			Set \(\eta_\ell = \eta_\ell+1\)\;
		}
		\KwOut{\(\appEub{0}{\ell+\eta_\ell}\p{\nvar{z}}\)}
		\caption{Adaptive sampling procedure at level \(\ell\)}
		\label{alg:adaptivity}
	\end{algorithm}
\end{center}

\subsubsection{Optimal Starting Level}\label{sec:practical_mlmc}
The proof of Theorem\nobreakspace \ref {thm:ad_mlmc_hierarchy} uses asymptotic results on \(\var{\dhl}\) and \(\E{N_{0,\ell+\eta_\ell}}\) to bound the cost of the MLMC estimator \eqref{eqn:cva_pll_mlmc} with adaptive differences \eqref{eqn:dhl_adapt}. In practice, it is possible that the differences \(\dhl\) exhibit pre-asymptotic behaviour at small levels \(\ell\approx 0\). In particular, defining \(\dhl^f\defeq \I{\appEub{0}{\ell+\eta_\ell}\p{\nvar{z}}>\lloss}\) to represent the finest approximation in \eqref{eqn:dhl_adapt}, it is important that the ratio
\[
	R_{\ell_0} \defeq \frac{\sqrt{\var{\dhl[\ell_0]^f}N_{0,\ell_0}} + \sqrt{\var{\dhl[\ell_0+1]}N_{0,\ell_0+1}}}{\sqrt{\var{\dhl[\ell_0+1]^f}N_{\ell_0+1}}},
\]
\nt{is less than 1}, otherwise using \eqref{eqn:mlmc_cost} we can see it is more efficient to start the MLMC estimate from level \(\ell_0+1\). In \cite{GilesHajiAli:2018}, the authors suggest choosing \(\ell_0\) a-priori such that \(R_{\ell_0}\le 1\). Note that for the adaptive refinements defined by Algorithm\nobreakspace \ref {alg:adaptivity}, increasing \(\ell_0\) is not the equivalent to increasing the base sample size, \(N_{0,0}\)\nt{, unlike a non-adaptive algorithm}.

\subsubsection{Robust Error Estimation}\label{sec:high_kurtosis}
Robust evaluation of the MLMC estimators \eqref{eqn:cva_pll_mlmc} requires accurate estimates of the error terms \(V_\ell \defeq \var{\dhl}\) and \(E_\ell \defeq \abs{\E{\dhl}}\) to approximate the optimal maximum level $L$ and number of samples $\Mvec$. Since \(\abs{\dhl}\) is a Bernoulli random variable, reliable Monte Carlo estimates for $V_\ell$ require $\Order {\kurt{\dhl}} = \Order{V_{\ell}^{-1}}$ samples, contradicting the fact that MLMC aims to reduce the number of samples required at {fine} levels and forcing a trade-off between efficiency and  robustness of the estimator.
In \cite{xu2020}, the authors approximate $\H{\cdot}$ using a sequence of smooth logistic sigmoid functions which depend on $\ell$ and are capable of reducing $\kurt{\dhl}$ by a constant factor. Alternatively, \final{Collier et al..} \cite{Collier:CMLMC} assume $V_\ell \approx a_02^{-\beta\ell}, E_\ell\approx a_12^{-\alpha\ell}$ for general MLMC problems and use Bayesian estimates for the parameters $a_0, a_1, \alpha, \beta$ based on the observed samples at levels \(\ell=0,\dots, L\). In \cite{Elfverson:2016selectiverefinement}, the authors consider a Bayesian approach which aims to estimate the probability that \(\dhl\) is non-zero. An extension of this Bayesian approach is outlined in Appendix\nobreakspace \ref {app:bayes_err}.

\section{Model for the Credit Valuation Adjustment}\label{sec:cva_cva}
We now apply the hierarchical estimator \eqref{eqn:cva_pll_mlmc} constructed in Section\nobreakspace \ref {sec:cva_mlmc} to the problem of computing the probability of large loss due to fluctuations in the CVA \eqref{eqn:cva_master}. Consider a financial market with \(d\) assets \(\p{S_t}_{0\le t\le T}\) with dynamics modelled by the Brownian SDE \eqref{eqn:cva_sde} under a physical measure \(\pmeasure\) and which are adapted to the filtration \(\filt[set]\). The discounting factor is modelled by the dynamics \(\text{d}B_t = r_tB_t\text{d}t\), with \(B_0 = 1\), where \nt{the process \(r\defeq\p{r_t}_{t\ge0}\) is adapted to the filtration \(\filt[set]\), with \(r_t\) taking values in \(\rset_+\) for all \(t\ge0\)}. To simplify the presentation, it is assumed that \(B_t\) can be sampled \st{exactly}. However, the slightly more general case where \(B_t\) requires approximate sampling and \(r_t\) follows a Brownian SDE can be tackled naturally by setting \nt{\(\tilde S_t\defeq \p{S_t, r_t, B_t}\) for all \(t\ge 0\)}.  We assume the existence of a risk-neutral measure, \(\qmeasure\), under which the assets evolve under the risk-neutral SDE
\begin{equation}\label{eqn:sde_rn}
	\text{d}S_t = r_tS_t\text{d}t + \sdevol\p{t, S_t}\text{d}W_t^{\qmeasure},
\end{equation}
where \(W^\qmeasure\defeq \p{W_t^\qmeasure}_{t\ge 0}\) is a \(d^\prime\)-dimensional \(\qmeasure\)-Brownian motion.

Consider a contract between a bank and a counterparty, which has a single payoff occurring at maturity \(T\), \st{depending only on \(S_T\),} denoted by \(\pi\p{S_T}\) for some \(\pi:\rset^d\to\rset\). \st{The risk-neutral value of the \st{bank's} portfolio at time \(t\in\sq{0,T}\) is hence \(\val{t}= B_t\E^\qmeasure{B_T^{-1}\pi\p{S_T}\given \filt} = B_t\E^\qmeasure{B_T^{-1}\pi\p{S_T}\given S_t}\). } We emphasise here that the restriction of the portfolio to a single payoff and counterparty is made solely with the intention of simplifying the following discussion and apply equally to any number of counterparties, see Section\nobreakspace \ref {sec:cva_multiple_counterparties}. \st{The restriction to a European payoff with a Markovian value process is made both for ease of notation in what follows and to simplify the numerical integration of the SDE \eqref{eqn:sde_rn} in accordance with Assumption \ref{assumpt:cva_general_rv_convergence} by removing the effects of non-Markovian dynamics on the numerical integration error. An extension to non-Markovian portfolios is discussed further in Remark \ref{rem:non_markovian}}. Counterparty credit risk is modelled through the stopping time \(\tau\) with respect to the expanded filtration \(\filt[set][G]\) that \nt{incorporates} additional credit risk factors.   To account for potential counterparty default, the value of the \st{bank's} position at time \(t\) is adjusted by the amount \(\cva{t}\) defined in \eqref{eqn:cva_master}, with discounted payoffs \(\pi_{\tau, T} = B_\tau B_T^{-1}\pi\p{S_T}\), given \(S_\tau\), for \(\tau\le T\).  \st{The adjusted value of the \st{bank's} position, accounting for counterparty credit risk, is}
\begin{equation}\label{eqn:ad_val}
	\adval{t} = \val{t}\I{\tau\ge t} - \cva{t}.
\end{equation}
For the purposes of this work, we assume that the bank itself carries no risk of default. In settings where the bank carries a non-negligible default risk, a debit valuation adjustment (DVA) should also be considered in the adjusted valuation \eqref{eqn:ad_val} \cite{Gregory2020}, which is defined in the same manner as CVA but with the roles of the bank and counterparty reversed and therefore the methods discussed herein are equally applicable to the DVA.
 
Random losses for the adjusted valuation occurring by the short risk horizon \(\riskhor\ll T\) can be expressed as 
\[
\adloss \nt{\defeq} \adloss[\tau<\riskhor] + \adloss[\riskhor\le\tau],
\]
where \(\adloss[\tau<\riskhor]\) and \(\adloss[\riskhor\le\tau]\) reflect separately the nature of losses depending on whether \(\tau\) occurs before or after \(\riskhor\). When \(\tau\) falls before \(\riskhor\), losses in the valuation are given by the difference in (adjusted) valuations at time \(0\) and time \(\tau\). If default occurs before \(\riskhor\), the losses can be expressed as
\[
\begin{aligned}
\adloss[\tau<\riskhor] &\nt{\defeq} \I{\tau < \riskhor}\p[\Big]{\adval{0} - B_\tau^{-1}\adval{\tau}},
\end{aligned}
\]
whereas random losses in the event \(\tau \ge \riskhor\) take the form
\[
\begin{aligned}
\adloss[\riskhor\le\tau] &\nt{\defeq} \I{\riskhor\le\tau}\p[\Big]{\adval{0} - B_\riskhor^{-1}\adval{\riskhor}}\\
&= \underbrace{\I{\riskhor\le \tau}\p{\val{0} - B_\riskhor^{-1}\val{\riskhor}}}_{\eqdef\adloss[\textnormal{market}]} + \underbrace{\I{\riskhor\le \tau}\p{B_\riskhor^{-1}\cva{\riskhor} - \cva{0}}}_{\eqdef\adloss[\textnormal{credit}]}.
\end{aligned}
\]
In the above expression, the component \(\adloss[\textnormal{market}]\) captures the market risk associated \ntg{with} fluctuations in the risk-neutral valuation of the \st{bank's} position due to random fluctuations in the assets by the risk horizon \(\riskhor\). \final{An overview of Monte Carlo estimators for market risk measures of such losses can be found in Glasserman \cite{glasserman2004}}. Hierarchical estimators for such problems have been developed in \cite{GilesHajiAli:2018,GilesHajiAli:2019sampling,Gordy:2010,Brodie:2011}. Credit risk factors are captured by terms \(\adloss[\textnormal{credit}]\) and \(\adloss[\tau<\riskhor]\)\nt{, each considering} a subtly different aspect of counterparty credit risk:
\begin{itemize}
	\item \(\adloss[\textnormal{credit}]\) captures the risk posed by fluctuations in the credit worthiness of the counterparty impacting the valuation adjustment over the interval \(\sq{0,\riskhor}\), \st{reducing} the adjusted value of the \st{bank's} position.
	\item \(\adloss[\tau<\riskhor]\) captures the risk posed by imminent counterparty default leading to losses before the risk horizon \(\riskhor\).
\end{itemize}
In practice, under the physical measure which captures the underlying risk factors, \(\prob{\tau<\riskhor}\approx0\)  since \(\riskhor\ll T\) and the physical probability that the counterparty will default over a small period of time is small for counterparties with reasonable credit worthiness.  On the other hand, \final{as observed by Hull et al..} \cite{hw05} default probabilities are often inflated under the risk-neutral measure when compared to the physical measure \cite[Chapter 12]{Gregory2020}. It is therefore more likely that small fluctuations in the credit worthiness of the counterparty will lead to significant losses in the adjusted valuation of the portfolio through the CVA, as captured by \(\adloss[\textnormal{credit}]\). Consequently, in what follows we assume that \(\prob{\tau<\riskhor} = 0\), and focus on the losses \(\adloss[\textnormal{credit}]\) which are then equal to \(B_\riskhor^{-1}\cva{\riskhor} - \cva{0}\) almost surely under \(\pmeasure\).
Our goal is then to construct an efficient estimator for the probability \(\pll\) \nt{in} \eqref{eqn:cva_master},
that the losses \(\adloss[\textnormal{credit}]\) exceed a given threshold \(\lloss\) at the risk horizon \(\riskhor\ll T\).

\subsection{Model for the Default Times}\label{sec:cva_default_times}
Following the approach in \cite[Section 3.3]{BrigoDamiano2013Ccrc} \nt{and} \cite[Chapter 12]{Gregory2020}, we assume an intensity model for the default times under the risk-neutral measure. Specifically, for a positive-valued \nt{process \(h = \p{h_t}_{t\ge 0}\), adapted to \(\filt[set][G]\), representing a hazard rate,} assume that under the risk-neutral measure
\(
	\qprob{\tau \le t} = 1 - \exp\p{-\int_{s=0}^t h_s\text{d}s}
\)
for all \(t\ge 0\). Then, the instantaneous probability of default at time \(t\) is \(h_t\text{d}t\). A common model for the hazard rates is
\(
	h_t = \lgd^{-1}\ct{t},
\)
which depends on the loss given default and the counterparties credit spread, \(\ct{t}\), which is a positive-valued, \nt{\(\filt[t][G]\)-measurable, random variable} representing the difference in return between a counterparty bond and the risk-neutral bond \(B_t\) \cite[Section 12.2.3]{Gregory2020}.  A large credit spreads implies the counterparty bond has high associated risk, which is a sign of poor credit worthiness and large default probabilities. Following \cite{Pykhtin2012}, we assume that the credit spread is constant under the risk-neutral measure \(\qmeasure\). Then, for any fixed \(v \st{>} 0\), we have:
\[
	\qprob{v\le \tau\le t+v\given \tau\ge v, \ct{v}} = 1 - \exp\p[\bigg]{-\frac{\ct{v} t}{\lgd}}.
\] 
In particular, conditioned on \st{the event \(\br{\tau\ge v}\)}, the random variable \(\p{\tau - v}\) follows an \(\textrm{Exponential}\p{\ct{v}/\lgd}\) distribution.

As discussed above, we assume that \(\prob{\tau<\riskhor} = 0\). Following \cite{Pykhtin2012},  to model random fluctuations in the creditworthiness of the counterparty over the short risk horizon, we assume that the credit spread follows a geometric Brownian motion under \(\pmeasure\). In particular, assume
\begin{equation} \label{eqn:cva_cspread_pdynamics}
	\text{d}\ct{t} = \scs \ct{t}\text{d}\wcs{t},
\end{equation}
where \(c_0>0\) can be inferred from market data, \(\scs\) is a constant representing the volatility of the credit spread and \(\br{\wcs{t}}_{0\le t\le T}\) is a \(\pmeasure\)-Brownian motion. Large values of the product \(\scs\ct{0}\) imply that there is high initial volatility in the credit spread. This can lead to significant worsening of the credit worthiness of the counterparty by the risk horizon \(\riskhor\), causing large losses due to fluctuations in the CVA. In \cite[Table A]{Pykhtin2012}, the product \(\scs\ct{0}\) is directly related to the credit rating of the counterparty to compute the standardised CVA formula.

To summarise, the above model for default times relies on two assumptions: Under the physical measure, the physical probability of counterparty default before \(\riskhor\) is negligible, but the credit spread fluctuates according to a geometric Brownian motion. As a result, risks posed by changing credit worthiness of the counterparty are captured by the dynamics \eqref{eqn:cva_cspread_pdynamics} of \(\ct{t}\) under \(\pmeasure\). On the other hand, under the risk-neutral measure, conditioned on the credit spread \(\ct{v}\) at time \(v\ge 0\) and the event \(\tau\ge v\), the credit spread is fixed and the random variable \(\p{\tau - v}\) follows an \(\textrm{Exponential}\p{\ct{v}/\lgd}\) distribution. Important to what follows, this model allows us to sample random default times \nt{\(\tau\) under the risk-neutral measure}, conditioned on a given risk scenario, \nt{exactly}. More general risk-neutral models for the default time are discussed in \cite{Gregory2020,BrigoDamiano2013Ccrc}, which do not rely on constant credit spreads under \(\qmeasure\). However, such models can lead to difficulties in sampling the default times which are beyond the scope of algorithms presented in this paper, and are left to future work. Nevertheless, as remarked in \cite{Pykhtin2012}, the assumption of constant credit spreads is practical since the information used to model credit spreads is often sparsely distributed in time. 

\subsection{Formulation as a Nested Expectation}\label{sec:cva_nest_exp}
Under the assumption that no default occurs under the physical measure and for the default time model in Section\nobreakspace \ref {sec:cva_default_times}, the (discounted) CVA at time \(t\) is equivalent to
\[
	\begin{aligned}
		\frac{\cva{t}}{B_t} = \E[\bigg]^{\qmeasure}{\I{\defc{t}\le T}\lgd \max\br[\Big]{\E[\big]^{\qmeasure}{B_T^{-1}\pi\p{S_T}\given \srt[t]{\defc{t}},\defc{t}}, 0}\given \srhor[t], \ct{t}},
	\end{aligned}
\]
where we introduce the following notation:
\begin{itemize}
	\item \(\srhor[t]\) is \nt{superscripted} by \(\pmeasure\) to indicate it is driven by the \(\pmeasure\)-Brownian motion \(W\) through the dynamics \eqref{eqn:cva_sde}.
	\item \(\defc{t}\) is indexed to highlight the \st{dependence} of the default times on the credit spread \(\ct{t}\). In the setup of Section\nobreakspace \ref {sec:cva_default_times}, \(\defc{t} - t\) is an \(\textrm{Exponential}\p{\ct{t}/\lgd}\) random variable. 
	\item For \(t\le u\le T\) and \(x\in\rset^d\), \(\srt[t][x]{u}\) represents the asset value \(S_u\) given \(\srt[t][x]{t} =x\).
\end{itemize} 
In particular, \final{\ntg{defining \(f\p{u} = {\max\br{u, 0}}\) for $u\in\rset$ and, by an abuse of notation, defining the function  for $s\in\rset^d$ and ${t_\text{def}>0}$ given by} \[{\nestE{1}}\p{s, t_\text{def}} = \lgd\I{t_\text{def}\le T}\E^{\qmeasure}{B_T^{-1}\pi\p{S_T}\given S_\tau=s, \tau=t_\text{def}},\]} we have
\[
	\frac{\cva{t}}{B_t} = \E[\Big]^{\qmeasure}{\fcva{t}\given \srhor[t], \ct{t}}.
\]
\final{By a further abuse of notation, for $s\in\rset^d$ and $c>0$} the problem \eqref{eqn:cva_master} becomes
\begin{equation}\label{eqn:cva_cva_var_model}
	\begin{aligned}
		\pll &=  \prob[\big]{{\nestE{0}}\p{\srhor, \ct{\riskhor}} > \lloss},\\
		{\nestE{0}}\nt{\p{s, c}} &\defeq \E[\Big]^{\qmeasure}{\fcva{\riskhor}\given \srhor\nt{=s}, \ct{\riskhor}\nt{=c}} - \cva{0},
	\end{aligned}
\end{equation}
\nt{where \(\srt[\riskhor]{\defc{\riskhor}}\) represents the risky asset values at default, first simulated under the physical measure up to the horizon \(\riskhor\), and then simulated up to the default time \(\defc{\riskhor}\) using the risk-neutral dynamics \eqref{eqn:sde_rn} driven by the \(\qmeasure\)-Brownian motion \(W^\qmeasure\).} In contrast, \(\srt[0]{\defc{0}}\) is simulated entirely under the measure \(\qmeasure\) up to the default time \(\defc{0}\), which depends only on the initial credit spread \(\ct{0}\).
This problem can be expressed in the general form \eqref{eqn:cva_general_nested_prob}, \nt{where}
\begin{equation}\label{eqn:cva_cva_var_link}
	\begin{aligned}
		\nvar{z} &= \p{\srhor, \ct{\riskhor}} \quad \text{\nt{is \(\filt[\riskhor][G]\)-measurable, taking values in \(\rset^{d+1}\)}}\\
		\nvar{y} &= \Big(\srt[\riskhor]{\defc{\riskhor}}, \defc{\riskhor}\Big) \quad \text{\nt{is \(\filt[\tau_{c_\riskhor}][G]\)-measurable, taking values in \(\rset^{d+1}\)}}\\
		\nvar{x} &= \ntg{\lgd} \I{\defc{\riskhor}\le T}B_T^{-1}\pi\Big(\srt[\riskhor]{T}\Big) \quad \text{\nt{is \(\filt[T][G]\)-measurable, taking values in \(\rset\)}}.
	\end{aligned}
\end{equation}
The variable \(\nvar{x}\) \nt{contains a discontinuous function of the default time \(\defc{\riskhor}\)}. In the current setting, this has no influence on the complexity of the MLMC estimation of \(\pll\) (equivalently \(\lloss\)) since the current framework samples the default times exactly. More generally, this discontinuity can be removed by using importance sampling techniques, as we will discuss in Section\nobreakspace \ref {sec:cva_importance_sampling}, to ensure the sampled default times always fall before \(T\).
Within the computation \eqref{eqn:cva_cva_var_model}, \(\cva{0}\) is independent of the generated risk scenario \(\srhor\) and  \(\ct{\riskhor}\) and can be approximated a-priori outside of the nested expectation over the risk factors. In Section\nobreakspace \ref {sec:cva_cva_capital}, we show that by including \(\cva{0}\) as part of the inner computation we can leverage dependencies between \(\srt[0]{t}\) and \(\srt[\riskhor]{t}\) to reduce the variance of the estimator. 

\st{
\begin{remark}
	\label{rem:non_markovian}
	To account for Non-Markovian portfolios, the random variables \(\nvar{z}\) and \(\nvar{y}\) in \eqref{eqn:cva_cva_var_link} can be replaced by the infinite dimensional processes \({\nvar{z} = \br{W_t}_{0\le t \le \riskhor} \cup \br{\ct{\riskhor}}}\) and \(\nvar{y}=\br{W_t^\qmeasure}_{\riskhor\le t\le \defc{\riskhor}} \cup \br{\defc{\riskhor}}\). Provided the (discrete) approximation of these quantities and the payoff \(\nvar{x}\) satisfies Assumption \ref{assumpt:cva_general_rv_convergence}, the methods in this paper can be adapted naturally to this setting. {The construction of MLMC methods satisfying Assumption\nobreakspace \ref{assumpt:cva_general_rv_convergence} in the case of portfolios containing Asian and lookback options can be found in the work of \final{Giles, Debrabant and R{\"o}ssler \cite{giles2013analysis}.} For portfolios where these convergence rates cannot be achieved due to irregularities in the payoff, for example for Barrier or Digital options \cite{giles2013analysis}, or non-Lipschitz market dynamics as considered by Derouich and Kebaier \cite{derouich2022interpolated}, the analysis in Section\nobreakspace \ref{sec:cva_unbiased_mlmc} can be adapted to include nested biased MLMC estimation, with less restricted assumptions on the convergence rates. Such an approach is beyond the scope of this paper, and can be found in \cite{spence23}.}
\end{remark}
}

\subsection{Collateralization}\label{sec:collateralisation}
A more realistic model for losses arising due to CVA should incorporate credit risk mitigation factors, with the aim of lowering the risk faced by institutions due to counterparty credit risk. In recent years, the initial and variational margins have become essential in reducing the exposure of the bank to its counterparties \cite{Gregory2020,GreenAndrew2015XCFa}. Assuming credit risk arises only through the counterparty, the variational margin requires the counterparty to post funds to a segregated account (free from default-risk) at a sequence of discrete time-points to cover the cost of their exposure to the bank. The initial margin is then based on a (conditional) Value-at-Risk formula on the exposure evaluated at each posting time of the variational margin to account for fluctuations which arise between posting times. Computing variational adjustments therefore requires evaluating additional nested expectations defining the exposure faced to the counterparty at the posting time immediately prior to default. On the other hand, initial margin computation requires a nested value-at-risk calculation, given the market state at the posting time prior to counterparty default. Defined in full generality, inclusion of nested risk measures will require, for example, nested stochastic approximation schemes to approximate the initial margin. Such methods fall beyond the scope of this paper and are left as an interesting avenue for future work. In the particular case where the assets \(S\) evolve according to a geometric Brownian motion, the initial margin can be related to the solution of a McKean backward stochastic differential equation as in Agarwal et al.. \cite{adgl19} and Henry-Labord\'ere \cite{hl17}. In \cite{Bourgey20}, the authors use this approach to co\text{d}mpute the initial margin by approximating a pair of nested expectations using an antithetic MLMC estimator as discussed in Section\nobreakspace \ref {sec:cva_hierarchical_mc}. Defined in this manner, the initial margin computation will add an extra layer of nested expectations when computing the CVA. \nt{The additional layer of nested expectations} can be handled by extending the estimator in Section\nobreakspace \ref {sec:cva_hierarchical_mc} to handle an additional nested expectation, which is possible using MLMC techniques for repeatedly nested expectations as discussed in \cite{spence23,zhou2022,syed2023optimal}.  
\section{Approximation of the Capital Charge}\label{sec:cva_cva_capital}
With the aim of reducing the variance of \({\nestE{0}}\p{\srhor, \ct{\riskhor}}\) in \eqref{eqn:cva_cva_var_model}, we include the computation of \(\cva{0}\) within the conditional average over \(\srhor, \ct{\riskhor}\) through
\[
	{\nestE{0}}\p{\srhor, \ct{\riskhor}} = \E[\Big]{\lcva\p[\big]{{\nestE{1}}\p{\srt{\defc{\riskhor}},\defc{\riskhor}}, {\nestE{1}}\p{\srt[0]{\defc{0}},\defc{0}}}\given \ct{\riskhor}, \srhor},
\]
where \nt{we define the function \(\lcva\p{u, v} \defeq f\p{u} - f\p{v}\) and, by an abuse of notation, refer to the random variable
\begin{equation}\label{eqn:cva_inner_samples}
\begin{aligned}
\lcva &\defeq \lcva\p[\big]{{\nestE{1}}\p{\srt{\defc{\riskhor}},\defc{\riskhor}}, {\nestE{1}}\p{\srt[0]{\defc{0}},\defc{0}}}.
\end{aligned}
\end{equation}}
The assets \(\srt{}\) and \(\srt[0]{}\) can be approximated under the risk-neutral measure \nt{\(\qmeasure\)} using \nt{the} Milstein scheme \cite{Kloeden:1999} and the methods of Section\nobreakspace \ref {sec:cva_mlmc} \nt{can be} applied directly, using the linearity of expectation to handle the difference between both averages \({\nestE{1}}\) within \eqref{eqn:cva_inner_samples}, to compute \(\pll\).
There are several aspects of the problem in this form which impede efficient multilevel estimation:
\begin{itemize}
	\item The variance of the inner samples \(\lcva\)
	within \eqref{eqn:cva_cva_var_model}, is inflated by the two different sources of risk:
	\begin{itemize}
		\item (Credit Risk). The default times \(\defc{\riskhor}\) and \(\defc{0}\) occur with different rates and it is difficult to correlate a pair of samples under each rate effectively. It is unlikely that independent samples of \(\defc{\riskhor}\) and \(\defc{0}\) will both occur before maturity \(T\).
		\item (Market Risk). The processes \(\srt{}\) and \(\srt[0]{}\) evolve under the measures \(\pmeasure\) and \(\qmeasure\), respectively, on the interval \(\sq{0,\riskhor}\). 
	\end{itemize}
	\item Under the risk-neutral measure, the events \(\defc{\riskhor}\le T\) and \(\defc{0}\le T\) are rare, leading to instability within the MLMC estimate of \(\pll\). 
\end{itemize}

In Section\nobreakspace \ref {sec:cva_var_reduct}, several variance reduction techniques are used to replace \(\lcva\) with a variable \(\lcva^{\textrm{III}}\) satisfying \(\E{\lcva^{\textrm{III}} \given \srhor, \ct{\riskhor}}  = \E{\lcva \given \srhor, \ct{\riskhor}}\), but with \(\var{\lcva^{\textrm{III}}\given \srhor, \ct{\riskhor}} = \Order{\riskhor}\), constituting a significant reduction in variance. In Section\nobreakspace \ref {sec:cva_importance_sampling}, importance sampling techniques are used to sample default times from the conditional distribution of the counterparty defaulting before \(T\), resolving instability issues from rare event sampling.  

\subsection{Variance Reduction}\label{sec:cva_var_reduct}
We begin by using a change of measure to remove the need to sample \(\defc{\riskhor}\) under a different rate from \(\defc{0}\). For the exponentially distributed default times in Section\nobreakspace \ref {sec:cva_default_times} and any \(F:\rset_+\to\rset\), we have
\[
\begin{aligned}
\E^{\qmeasure}{F(\defc{\riskhor}) |\ct{\riskhor}} &= \int_\riskhor^\infty F(t)\frac{\ct{\riskhor}}{\lgd}e^{-\ct{\riskhor}\p{t-\riskhor}/\lgd}\text{d}t\\
&= \frac{\ct{\riskhor}}{c_0}e^{\ct{\riskhor} \riskhor/\lgd}\int_0^\infty \I{t\ge \riskhor} F\p{t} \frac{c_0}{\lgd} e^{-\p{\ct{\riskhor} - c_0}t/\lgd} e^{-c_0 t/\lgd}\text{d}t\\
&= \E[\big]^{\qmeasure}{\I{\defc{0}\ge \riskhor}g_{\defc{0}}(\ct{\riskhor})e^{\ct{\riskhor} \riskhor/\lgd}F\p{\defc{0}}\given\ct{\riskhor}},
\end{aligned}
\]
where we define $g_t(x) \defeq c_0^{-1}xe^{-\p{x-c_0}t/\lgd}$. In particular,
\[
	\begin{aligned}
	&\E[\Big]^{\qmeasure}{\fcva{\riskhor}\given \srhor,\ct{\riskhor}}\\
	& = \E[\Big]^{\qmeasure}{\I{\defc{0}\ge \riskhor}g_{\defc{0}}(\ct{\riskhor})e^{\ct{\riskhor} \riskhor/\lgd}\fcontrolvar \given \srhor,\ct{\riskhor}},
	\end{aligned}
\]
where the change of measure moves the \st{dependence} on \(\ct{\riskhor}\) to  the \st{random variable} \(g_{\defc{0}}(\ct{\riskhor})e^{\ct{\riskhor} \riskhor/\lgd}\), \st{simulating the} default \nt{time} using \(\defc{0}\) opposed to \(\defc{\riskhor}\), eliminating the need to sample \(\defc{\riskhor}\). 
Consequently, we henceforth express \(\defc{0}\equiv \tau\).
The above discussion then motivates replacing \(\lcva \) as in \eqref{eqn:cva_inner_samples} with
\begin{equation}
\label{eqn:lambd1}
\begin{aligned}
\lcva^{\textrm{I}}
\defeq
\I{\tau\ge \riskhor}\bigg(g_\tau(\ct{\riskhor})&e^{\ct{\riskhor} \riskhor/\lgd}\underbrace{\p[\Big]{\fcontrolvars - \fcvas{0}}}_{\text{Market risk}} \\
& +\underbrace{\p[\big]{g_\tau\p{\ct{\riskhor}}e^{\ct{\riskhor} \riskhor/\lgd} - 1}}_{\text{Credit risk}} \fcvas{0} \bigg), 
\end{aligned}
\end{equation}
whenever \( \tau \ge \riskhor\). By construction \(\E^{\qmeasure}{\lcva^{\textrm{I}}\given\srhor,\ct{\riskhor}} = \E^{\qmeasure}{\I{\defc{0}\ge\riskhor}\lcva\given\srhor,\ct{\riskhor}}\). The component of $\lcva$ reflecting default before $\riskhor$\nt{, under \(\qmeasure\),} is
\[
\I{\tau<\riskhor}\lcva =  -\I{\tau<\riskhor}\fcvas{0}.
\]
Since this is independent of the risk factors {at the horizon \(\riskhor\)}, we can cancel the bias induced by considering only $\tau\ge \riskhor$ in $\lcva^{\textrm{I}}$ by setting
\[
\lloss^{\textrm{I}} \defeq \lloss + \E[\Big]^{\qmeasure}{\I{\tau< \riskhor}\fcvas{0}},
\] 
which be pre-computed offline at insignificant cost compared with the cost of computing $\pll$.
By the change of measure from \(\defc{\riskhor}\) to \(\tau\) above, it follows that $\E^{\qmeasure}{\lcva - \lloss|\srhor, \ct{\riskhor}} = \E^{\qmeasure}{\lcva^{\textrm{I}} - \lloss^{\textrm{I}}|\srhor, \ct{\riskhor}}$. We consider variance reduction for the credit and market risk factors in turn.

\subsubsection{Credit Risk}
\label{sec:var-credit}
In \cite{Pykhtin2012}, it is shown that the standardised CVA capital charge can be derived by linearising  the term \(g_\tau\p{\ct{\riskhor}}e^{\ct{\riskhor}} \) as a function of \(\ct{\riskhor}\). We use this approach here to construct a control variate \nt{that reduces the} variance of \(\lcva\). In particular, since \(g_\tau\p{\ct{0}} = 1\), we have
\begin{equation*}
\label{eqn:gtau_lin}
\begin{aligned}
g_\tau\p{\ct{\riskhor}}e^{\ct{\riskhor} \riskhor/\lgd} - 1&= g_\tau\p{\ct{\riskhor}}\p[\big]{1 + \Order{\riskhor}} - g_\tau\p{\ct{0}} \\
&= \underbrace{\p{\ct{\riskhor} - c_0}\p[\bigg]{\frac{1}{c_0} - \frac{\tau}{\lgd}}}_{\Order{\sqrt{\riskhor}}} +\Order{\riskhor},
\end{aligned}
\end{equation*}
where \nt{the} difference \(\ct{\riskhor} - c_0\) is of order \(\sqrt{\riskhor}\) since the process \nt{\(\p{\ct{t}}_{t\ge 0}\)} evolves under \(\pmeasure\) according to the dynamics \eqref{eqn:cva_cspread_pdynamics} with an underlying \(\pmeasure\)-Brownian motion.
To remove this leading order term, we set
\[
\lcva^{\textrm{II}} \defeq \lcva^{\textrm{I}} - \I{\tau\ge \riskhor} \p{\ct{\riskhor} - c_0}\defcv_\tau\p{\srt[0]{\tau}},
\]
where we introduce the control variate 
\[
\defcv_\tau\p{\nt{s}} \defeq \p[\bigg]{\frac{1}{c_0} - \frac{\tau}{\lgd}}f\p[\big]{{\nestE{1}}\p{s, \tau}}.
\]
To ensure the control variate introduces no bias, we set
\[
\lloss^{\textrm{II}} \defeq \lloss^{\textrm{I}} - \p{\ct{\riskhor}-c_0}\E[\big]^{\qmeasure}{\defcv_\tau\p{\srt[0]{\tau}}},
\]
so that again $\E^{\qmeasure}{\lcva^{\textrm{II}} - \lloss^{\textrm{II}}|\srhor, \ct{\riskhor}} = \E^{\qmeasure}{\lcva - \lloss|\srhor, \ct{\riskhor}}$. \nt{Since \(\defcv_\tau\p{\srt[0]{\tau}}\) does not depend on the risk factors,
\(
	\E^{\qmeasure}{\defcv_\tau\p{\srt[0]{\tau}}} 
\) 
can be computed offline at negligible cost. Consequently, \(\lloss^{\textrm{II}}\) may be computed with bounded cost for each risk scenario \(\ct{\riskhor}\).}
\subsubsection{Market Risk}
\label{sec:var-reduct-market}
The market risk component of \(\lcva\) is primarily composed of the term
\[
 \fcontrolvars - \fcvas{0} = \fb\p{\srt{\tau}} - \fb\p{\srt[0]{\tau}},
\]
where we introduce the functions \(\fb:\rset^d\to\rset\), dropping the dependency of \(\fb\) on \(\tau\), to simplify notation in what follows. As with the credit risk, we focus on the case where \(\riskhor\le \tau\le T\).
Following the techniques in \cite[Section 3.2]{GilesHajiAli:2019sampling}, this term can be decomposed into
\[
	\begin{aligned}
		&\fb\p{\srt{\tau}} - \fb\p{\srt[0]{\tau}}\\
		&= \p[\big]{\fb\p{\srt{\tau}} - \fb\p{\srtt{\tau}{0}}} + \p[\big]{\fb\p{\srtt{\tau}{0}} - \fb\p{\srt[0]{\tau}}}\\
		&= \nt{\p[\big]{\srt{\tau} - \srtt{\tau}{0}}^\intercal\nabla \fb\p{\srtt{\tau}{0}}} + \p{\srhor - \srhor[0]}^\intercal \nabla_{\srhor[0]} \fb\p{\srt[0]{\tau}} + \Order{\riskhor},
	\end{aligned}
\]
using a Taylor expansion in the final step. Both of the terms \(\srt{\tau} - \srtt{\tau}{0}\) and \(\srhor - \srhor[0]\) are of order \(\sqrt{\riskhor}\) since they involve simulating the assets under different Brownian motions along the interval \(\sq{0,\riskhor}\). The former term can eliminated by \nt{utilising} an antithetic pair of risk-neutral paths \(\srtpm{}{\pm}\), each of which follows the \nt{\(\qmeasure\)-Brownian path} \(\br{\pm W_t^{\qmeasure}}_{0\le t\le \riskhor}\) up to the risk horizon. After the risk horizon, each of the processes \(\srt{}\) and \(\srtpm{}{\pm}\) should follow the same Brownian path \(\br{W_t^{\qmeasure}}_{\riskhor\le t \le \tau}\) up to the default time. It then follows for the antithetic variate \cite{GilesHajiAli:2019sampling,glasserman2004}
\[
	\begin{aligned}
	&\fb\p{\srt{\tau}} - \frac{1}{2}\p[\big]{\fb\p{\srtpm{\tau}{+}} + \fb\p{\srtpm{\tau}{-}}}\\
	&= \frac{1}{2}\p{\srhor - \srhor[0]}^\intercal\p[\big]{\nabla_{\srhor[0]}  \fb\p{\srtpm{\tau}{+}}+\nabla_{\srhor[0]}  \fb\p{\srtpm{\tau}{-}}} + \Order{\riskhor}.
	\end{aligned}
\] 
By subtracting the term
\[
	\frac{1}{2}\p{\srhor - \srhor[0]}^\intercal\p[\big]{\nabla_{\srhor[0]} \fb\p{\srtpm{\tau}{+}}+\nabla_{\srhor[0]}  \fb\p{\srtpm{\tau}{-}}},
\]
one obtains samples which are of \(\Order{\riskhor}\), constituting a significant improvement to the variance. The components \(\nabla_{\srhor[0]} \cdot \fb\) are referred to in financial settings as the delta control variate \cite{glasserman2004}.

Following the above approach and using the decomposition in \eqref{eqn:lambd1} motivates the term
\[
\st{\begin{aligned}
\lcva^{\textrm{III}} \defeq \I{\tau\ge \riskhor}&\bigg(g_\tau\p{\ct{\riskhor}}e^{\ct{\riskhor} \riskhor/\lgd}f\p[\big]{{\nestE{1}}\p{\srt[\riskhor]{\tau}, \tau}} \\
&- \frac{1}{2}\p[\Big]{f\p[\big]{{\nestE{1}}\p{\srtpm[0]{\tau}{+}, \tau}} + f\p[\big]{{\nestE{1}}\p{\srtpm[0]{\tau}{-}, \tau}}}\\
&-\frac{1}{2} \p{\ct{\riskhor}-c_0}\p[\Big]{\defcv_\tau\p[\big]{\nt{\srtpm{\tau}{+}}} + \defcv_\tau\p[\big]{\nt{\srtpm{\tau}{-}}}}\\
& - \frac{1}{2}\p{S_\riskhor^{\riskhor, S_\riskhor^\pmeasure} - S_0}^\intercal\p[\Big]{\nabla_{S_0} f\p[\big]{{\nestE{1}}\p{\srtpm[0]{\tau}{+}, \tau}} + \nabla_{S_0} f\p[\big]{{\nestE{1}}\p{\srtpm[0]{\tau}{-}, \tau}}}\bigg),
\end{aligned}}
\]
As before, to compensate for the bias induced by the delta control variate, we replace $\lloss^{\textrm{II}}$ with
\[
\lloss^{\textrm{III}} \defeq \lloss^{\textrm{II}} -  \p{\srhor - S_0}^\intercal\E[\Big]^{\qmeasure}{\nabla_{S_0} \fcvas{0}}
\]
so that $\E^{\qmeasure}{\lcva^{\textrm{III}}\given\srhor, \ct{\riskhor}} - \lloss^{\textrm{III}} = \E^{\qmeasure}{\lcva\given\srhor, \ct{\riskhor}} - \lloss$. Since all of the expectations appearing in the definition of $\lloss^{\textrm{III}}$ are independent of $\srhor$ {and} $\ct{\riskhor}$, they may be pre-computed offline at insignificant cost.

\subsection{Importance Sampling}\label{sec:cva_importance_sampling}
Observe that the CVA loss, $\lcva^{\textrm{III}}$, is non-zero only when $\riskhor\le\tau\le T$. For credible counterparties this is a rare event, making it costly to simulate a single instance of default. To address this issue note that, defining $\tilde p \defeq \prob{\riskhor\le\tau\le T}$, we have
\begin{equation*}
\label{equation:importsampling}
\E^{\qmeasure}{\lcva^{\textrm{III}}\given \srhor, c_\riskhor} - \lloss^{\textrm{III}} = \E^{\qmeasure}{\lcva^{\textrm{III}}\tilde p \given \srhor, \ct{\riskhor}, \riskhor\le\tau\le T} - \lloss^{\textrm{III}}. 
\end{equation*}
Under the model for default times in Section\nobreakspace \ref {sec:cva_default_times}, \(\tau\) is an \(\textrm{Exponential}\p{\ct{0}/\lgd}\) random variable under the risk-neutral measure \(\qmeasure\). Hence,
for $\riskhor\le t \le T$, we have 
\[
\qprob{\tau \le t|\riskhor\le\tau\le T} = \frac{\qprob{\riskhor\le\tau\le t}}{\qprob{\riskhor\le\tau\le T}} = \frac{e^{-c_0\riskhor/\lgd} - e^{-c_0t/\lgd}}{e^{-c_0\riskhor/\lgd} - e^{-c_0T/\lgd}}.
\]
By inverting this equation for \(t\), we can \st{sample} from the conditional distribution using inverse transform sampling. This reduces the kurtosis of samples from the CVA loss by ensuring default always occurs.

The above approach relies explicitly on the model for default times described in Section\nobreakspace \ref {sec:cva_default_times}. More complex models of the default time will require more advanced importance sampling techniques to achieve a similar result. 

\subsection{Hierarchical MLMC Estimates of the Loss}\label{sec:cva_mlmc_cva}
\nt{To summarize}, combining the techniques above, we have expressed the problem \eqref{eqn:cva_cva_var_model} in the form
\[
	\begin{aligned}
		\pll &= \prob[\big]{\E^{\qmeasure}{\lcva \given\srhor, \ct{\riskhor}} > \lloss}\\
			&= \prob[\big]{\E^{\qmeasure}{\lcva^{\textrm{III}}\tilde p - \lloss^\textrm{III}\given\srhor,\ct{\riskhor}, \riskhor\le\tau\le T}>0},
	\end{aligned}
\]
where \(\lcva^{\textrm{III}}\) depends only on the default times \(\tau = \defc{0}\) and has significantly reduced variance compared with \(\lcva\) as in \eqref{eqn:cva_inner_samples}. Furthermore, \st{we are able to sample default times directly from the conditional distribution \(\tau\in\sq{\riskhor, T}\) using elementary importance sampling techniques}. To counteract the bias induced replacing \(\lcva\) with \(\lcva^{\textrm{III}}\), we set
\[
\begin{aligned}
	\lloss^{\textrm{III}} = \lloss &+ \E[\Big]^{\qmeasure}{\I{\tau< \riskhor}\fcvas{0}}\\
								   &- \p{\ct{\riskhor}-c_0}\E^{\qmeasure}{\defcv_\tau\p{\srt[0]{\tau}}}\\
								   &- \p{\srhor - S_0}^\intercal\E[\Big]^{\qmeasure}{\nabla_{S_0} \fcvas{0}},
\end{aligned} 
\]
where each expectation can be pre-computed offline at insignificant cost using, for example, (multilevel) Monte Carlo sampling, since they do not depend on the generated risk variables. Since the variable \(\lcva^{\textrm{III}}\) is a functions of the nested expectations \({\nestE{1}}{\p{s,\tau}}\) for \(s \in\br{\srtpm[0]{\tau}{\pm}, \srt{\tau}}\), the techniques of Section\nobreakspace \ref {sec:cva_unbiased_mlmc}  can be used to construct antithetic correction terms \(\dl^{\p{\textnormal{ant}}}\lcva^{\textrm{III}}\) satisfying 
\[
	\E^{\qmeasure}{\lcva^{\textrm{III}}\tilde p - \lloss^\textrm{III}\given\srhor,\ct{\riskhor}, \riskhor\le\tau\le T} = \E[\bigg]^{\qmeasure}{\sum_{\ell=0}^\infty \dl^{\p{\textnormal{ant}}}\lcva^{\textrm{III}}\given\srhor,\ct{\riskhor}, \riskhor\le\tau\le T}.
\]
The antithetic differences \(\dl^{\p{\textnormal{ant}}}\lcva^{\textrm{III}}\) combine nested (unbiased) MLMC estimation of the expectations \({\nestE{1}}{\p{s,\tau}}\) with (Euler-Maruyama or Milstein) approximations of the underlying asset processes. Note that for the delta control variates used in \(\lcva^{\textrm{III}}\), we have
\[
	\nabla_{S_0} f\p[\big]{{\nestE{1}}\p{\srtpm[0]{\tau}{\pm}, \tau}} = \I{\tau\le T}\lgd \p[\Big]{\nabla_{S_0}\max\br[\big]{{\nestE{1}}\p{\srtpm{\tau}{\pm},\tau}, 0}},
\]
which is a discontinuous function of the nested expectation \({\nestE{1}}\p{\srtpm{\tau}{\pm}, \tau}\). Consequently, the delta control variates do not satisfy the conditions of \MakeUppercase assumption\nobreakspace \ref {assumpt:rec_f_decomposition}, \nt{and the analysis in} Section\nobreakspace \ref {sec:cva_unbiased_mlmc} \nt{cannot be applied directly}. One method of balancing the variance reduction properties of the delta control variate with the faster convergence of antithetic MLMC for Lipschitz continuous observables is to consider the delta control variates at a single level \cite{GilesHajiAli:2019sampling}. Defining
\[
		\tilde\lcva^{\textrm{III}} = \lcva^{\textrm{III}} + \frac{1}{2}\p[\Big]{S_\riskhor^{\riskhor, S_\riskhor^\pmeasure} - S_0}^\intercal\p[\Big]{\nabla_{S_0} \fcvaspm{0}{+} + \nabla_{S_0} \fcvaspm{0}{-}},
\] 
to be equivalent to \(\lcva^{\textrm{III}}\) but without the delta control variates, this motivates considering
\[
	{\tilde\Delta}_\ell^{\p{\textnormal{ant}}}\lcva^{\textrm{III}} 
	=
	\begin{cases}
	 \dl[0]\lcva^{\textrm{III}} & \ell = 0\\
	 \dl\tilde \lcva^{\textrm{III}} & \ell > 0.
	\end{cases}
\]
To remove the bias induced from including the delta control variates only at level zero, one should ensure the approximation of \(\E^{\qmeasure}{\nabla_{S_0} \fcvas{0}}\) within \(\lloss^\textrm{III}\) only contains a level zero of the approximation of \(\srt[0]{\tau}\) and \({\nestE{1}}\p{\cdot}\). Then 
\[
\E[\Big]^{\qmeasure}{\lcva^{\textrm{III}}\tilde p - \lloss^\textrm{III}\given\srhor,\ct{\riskhor}, \riskhor\le\tau\le T} = \E[\bigg]^{\qmeasure}{\sum_{\ell=0}^\infty {\tilde\Delta}_\ell^{\p{\textnormal{ant}}}\lcva^{\textrm{III}} \given\srhor,\ct{\riskhor}, \riskhor\le\tau\le T},
\]
can be used as in Section\nobreakspace \ref {sec:cva_mlmc} to construct an (adaptive) MLMC estimate of \(\pll\). Since Section\nobreakspace \ref {sec:cva_mlmc} assumed that the variable \(\nvar{z}\) in \eqref{eqn:cva_general_nested_prob} \nt{can be sampled \st{exactly}}, we assume that \(\srhor\) can be sampled exactly, which is a reasonable assumption provided \(\riskhor\ll T\). If \(\riskhor\) represents a significant time interval for which approximation of \(\srhor\) is costly, one can \nt{incorporate} multilevel approximations of \(\nvar{z}\defeq \srhor\) within the methods of Section\nobreakspace \ref {sec:cva_mlmc}. In this case, \MakeUppercase assumptions\nobreakspace \ref {assumpt:cva_general_rv_convergence} and\nobreakspace  \ref {assumpt:admlmc} should be extended naturally to include approximations \(\nvar[\ell]{z}\approx \nvar{z}\).

\subsection{Extension to Multiple Counterparties}\label{sec:cva_multiple_counterparties}
To simplify the preceding discussion, we have considered only the case where the contract occurs between the bank and a single counterparty.
Typically, the CVA capital charge considers large portfolios containing many options with \(\ncp\ge1\) counterparties. In this case, the adjustment made to the valuation of the \st{bank's} portfolio is 
\[
\cva{t} \defeq \sum_{n=1}^\ncp\cva{t}^{(n)},
\]
where \(\cva{t}^{(n)}\) denotes the valuation adjustment accounting for the \st{bank's} exposure to counterparty \(n\). In this approach, the credit spread \nt{at time \(t\ge 0\) becomes} \(\ct{t} = \p{\ct{t}^{(n)}}_{n=1}^\ncp\) and loss given default will differ for each counterparty.  The CVA capital charge becomes the value \(\lloss\) solving 
\begin{equation}
\label{eqn:cvavar-multcp}
\begin{aligned}
	\pll &= \prob[\bigg]{\sum_{n=0}^{\ncp}\p[\Big]{B_\riskhor^{-1}\cva{\riskhor}^{(n)} - \cva{0}^{(n)}} > \lloss}\\
		&= \prob[\bigg]{\sum_{n=0}^{\ncp}\E[\Big]^{\qmeasure}{\lcva^{\p{n}}\given \srhor, \ct{0}^{(n)}, \ct{\riskhor}^{(n)}} > \lloss}
\end{aligned}
\end{equation}
where \(\lcva^{(n)}\) is an extension of \eqref{eqn:cva_inner_samples} for the counterparty \(n\). While we focused on the case \(\ncp=1\) until now, the discussion applies in the extended case where \(\ncp>1\). We conclude this section by considering some additional points worth considering when \(\ncp\gg 1\).
\begin{itemize}
	\item (\textit{Randomized subsampling of counterparties.}) For a large number of counterparties (\(\ncp\gg 1 \) in \eqref{eqn:cvavar-multcp}) it can be more efficient to write the sum 
	\[
	\sum_{n=1}^\ncp \E^{\qmeasure}{\lcva^{\p{n}}\given \srhor, \ct{0}^{(n)}, \ct{\riskhor}^{(n)}} = \E^{\qmeasure}{\lcva^{\p{\kappa}}q_\kappa^{-1}\given \srhor, \ct{0}^{(\kappa)}, \ct{\riskhor}^{(\kappa)}},
	\]
	where \(\kappa\) is drawn from a discrete probability distribution on \(\{1, \dots, \ncp \} \) with mass function \(q_k\). As a result, we need only consider a single counterparty with each sample of the inner expectation rather than evaluate each of \(\{\lcva^{(n)} \}_{n=1}^\ncp\) for each sample. This approach, known as randomised sub-sampling \cite{GilesHajiAli:2019sampling}, can reduce the cost of the MLMC estimator \nt{for large values of \(\ncp\)}.
	
	\item (\textit{Simulation of market at large number of intermediate points.}) When there are many counterparties with different default times and/or many options per counterparty with differing maturities, it may be necessary to sample the assets at many intermediate time-points. In this case, it may be beneficial to use a linear interpolant to sample the market over discrete trajectories of the asset prices, opposed to generating discrete samples of the market at each default time. By combining the interpolant with an antithetic Milstein scheme, as in \cite[Theorem 4.13]{giles14antmilstein}, one may do this without affecting the convergence rates required within \MakeUppercase assumption\nobreakspace \ref {assumpt:cva_general_rv_convergence}.
	
\end{itemize} 
\section{Numerical Experiments}\label{sec:cva_num}
To illustrate the preceding methods, we consider the estimation of the probability of a large loss, $\pll$ \nt{in} \eqref{eqn:cva_cva_var_model}, for a model CVA portfolio consisting of a single stock process and \nt{a} single counterparty outlined in Section\nobreakspace \ref {sec:toyprob}. \st{The experiments in this section are performed using Python, the accompanying code may be found at \url{https://github.com/JSpence97/mlcva}.}

\subsection{Simple Synthetic Portfolio}
\label{sec:toyprob}

The underlying contract between the bank and counterparty is given by a linear combination of two European call options with \ntg{payoff} 
\[
\pi\p{\nt{s}} = {{c}_0\p{\nt{s} - K_0}^+ + {c}_1\p{\nt{s} - K_1}^+},
\]
where \nt{\(K_0,K_1\) are the strike price and} \nt{the {constants} ${c}_0$ and  ${c}_1$ are determined by a pre-defined value of \(B_0\E^\qmeasure{B_T^{-1}\pi\p{S_T}} \st{= v_0}\) and the delta-neutral condition \st{requiring} $\partial_{S_0}\E^\qmeasure{B_T^{-1}\pi\p{S_T}} = 0$. In particular, the delta-neutral condition implies that \(c_0\) and \(c_1\) are of opposite sign. Importantly, this ensures that the payoff \(\pi\) can be negative and thus the exposure to the counterparty can be zero, which retains the need to compute a nested expectation to approximate the CVA.}

We assume there is a single asset following a Geometric Brownian Motion under the Black-Scholes market assumptions with fixed interest rate $r = 1\%$ so that $B_t = e^{t/100}$. Then, under $\pmeasure$ and $\qmeasure$, respectively, \nt{the dynamics of the assets \(S\) are}
\begin{equation}\label{eqn:sde_gbm}
\begin{aligned}
\text{d}S_t &= \mu S_t\text{d}t + \volcva S_t\text{d}W_t^\pmeasure,\\
\text{d}S_t &= rS_t\text{d}t + \volcva S_t \text{d}W_t^\qmeasure.
\end{aligned}
\end{equation}
We take $T  = 1$ year, as well as $S_0 =  1$ and $\mu=\volcva = 0.1$. To test the methods in greater generality, we ignore the existing analytic formulae for the value and stock processes, resorting instead to Milstein simulation of \(S\) and Monte Carlo estimation of the exposure at default. Specifically, we use the \nt{nested} unbiased MLMC estimators \eqref{eqn:cva_pll_mlmc} to construct an estimator of \(\pll\) as in Section\nobreakspace \ref {sec:cva_mlmc}. 

Following the model for default times in Section\nobreakspace \ref {sec:cva_default_times}, we take $\ct{0} = 5\%$ and $\scs = 0.8\%/\ct{0}$, corresponding to an A-rated counterparty according to \cite[Table A]{Pykhtin2012}. The probability of large loss \eqref{eqn:cva_cva_var_model} is computed with risk horizon \(\riskhor = 10\) days and a fixed loss threshold \(\lloss = 5\times 10^{-4}\), giving \(\pll \approx 2\%\). 

\subsection{MLMC Setup}
\label{sec:mlmcset}
Simulation of the outer expectation in \eqref{eqn:cva_cva_var_model} \nt{under \(\pmeasure\)} is performed using the methods in Section\nobreakspace \ref {sec:cva_unbiased_mlmc}. In particular, we obtain the MLMC estimate \eqref{eqn:cva_pll_mlmc} which uses the multilevel correction terms
\[
\dhl = 	\I{\appEub{0}{\ell+\eta_\ell}\p{\srhor, \ct{\riskhor}}>\lloss} - \I{\appEub{0}{\ell-1+\eta_{\ell-1}}\p{\srhor, \ct{\riskhor}}>\lloss},
\]
where \({\nestE{0}}\p{{\srhor, \ct{\riskhor}}}\), defined by \eqref{eqn:cva_cva_var_model}, is approximated by the unbiased estimates \(\appEub{0}{\ell}\) defined through \eqref{eqn:cva_umlmc_middle} using \(\Mmell\) inner unbiased MLMC samples and random levels scaled by the parameters \(\zeta_0\) and \(\zeta_1\) as in Proposition\nobreakspace \ref {prop:cva_umlmc_moments}.
We use the techniques summarised in Section\nobreakspace \ref {sec:cva_mlmc_cva} to reduce the variance of samples  within \({\nestE{0}}\). To compute the probability, we use adaptive sampling  with \(\eta_\ell\) chosen according to Algorithm\nobreakspace \ref {alg:adaptivity} with \(\Mmell = \Mmell[0]2^\ell\) samples per level, where the refinement parameter \(\refp\) is chosen to be \nt{1.95} and the refinements use a normalising factor \(\sigma_\ell\) given by the corresponding conditional sample standard deviation  \eqref{eqn:sigma_sd}. To illustrate the effectiveness of Algorithm\nobreakspace \ref {alg:adaptivity}, we further consider non-adaptive sampling, taking \(\eta_\ell\equiv 0\), with \(\Mmell = \Mmell[0]2^{\gamma\ell}\) for \(\gamma = 1, 2\). At level \(\ell = 0\) we use \(\Mmell[0] = 8\) samples per level. \nt{Under} \MakeUppercase assumptions\nobreakspace \ref {assumpt:cva_general_rv_convergence} and\nobreakspace  \ref {assumpt:admlmc}, \nt{which hold for Milstein discretisation of the SDE \eqref{eqn:sde_gbm} by \cite[Theorem 10.3.5]{Kloeden:1999}}, and provided the condition \eqref{eqn:snn_cond} holds for the variables in \eqref{eqn:cva_cva_var_link}, it follows from Theorem\nobreakspace \ref {thm:ad_mlmc_hierarchy} and Remark\nobreakspace \ref {rem:thm_p_inf} that   \(\var{\dhl}\) is of order \(\Mmell^{-1/2 + \delta}\) and order \(\Mmell^{-1+\delta}\) for any \(\delta>0\) \nt{when} using non-adaptive and adaptive sampling, respectively.

For each MLMC estimate, we approximate the optimal starting level of MLMC using the iterative procedure in Section\nobreakspace \ref {sec:practical_mlmc}. Furthermore, as discussed in Section\nobreakspace \ref {sec:high_kurtosis}, MLMC level estimates of the probability \(\pll\) exhibit a high kurtosis. To obtain more stable \nt{variance and error} estimates, we use a biased Bayesian approach discussed in Appendix\nobreakspace \ref {app:bayes_err}.

\subsection{Results}
Figure\nobreakspace \ref {fig:cva_level-plots} plots various statistics of the multilevel correction terms \(\dhl\) for each method.
Figure\nobreakspace \ref {fig:cva_complexity} further plots the total work required to compute \(\pll\) to accuracy \(\tol\), against \(\tol\) for each method. We discuss the results below:
The top left plot in Figure\nobreakspace \ref {fig:cva_level-plots} shows the expected number of Gaussian and exponential random variables required to generate a single sample of \(\dhl\). By construction, since the cost of sampling the unbiased MLMC difference \eqref{eqn:delta_unbiased} is bounded by Proposition\nobreakspace \ref {prop:cva_umlmc_moments}, the non-adaptive samplers have \(\Order{2^{\gamma\ell}}\) cost (with some variability \ntg{due} to the random cost of the unbiased MLMC estimators). On the other hand, the adaptive sampler has expected cost \(\E{2^{\ell + \eta_\ell}} = \Order{2^\ell}\) as expected from Theorem\nobreakspace \ref {thm:ad_mlmc_hierarchy}.
The top right plot in Figure\nobreakspace \ref {fig:cva_level-plots} shows the variance \(V_\ell = \var{\dhl}\) as a function of \(\ell\). In agreement with taking the limit \(p\to\infty\) in \MakeUppercase assumption\nobreakspace \ref {assumpt:admlmc} and\nobreakspace Theorem\nobreakspace \ref {thm:ad_mlmc_hierarchy}, the non-adaptive samplers have \(V_\ell = \Order{N_{0,\ell}^{-1/2}} = \Order{2^{-\gamma\ell /2}}\). On the other hand, the adaptive sampler has \(V_\ell = \Order{N_{0,\ell}^{-1}} = \Order{2^{-\ell}}\). Combined with an expected sampling cost of order \(2^\ell\), this improves the convergence rate of the adaptive MLMC over the non-adaptive estimators. 
The bottom left plot in Figure\nobreakspace \ref {fig:cva_level-plots} shows the kurtosis of \(\dhl\) against \(\ell\). We observe \(\kurt{\dhl} = \Order{V_\ell^{-1}} \) for all methods. As mentioned above, this reduces the accuracy of standard estimates for the overall error made by MLMC, especially when \(L\) is large. This highlights the necessity to use \nt{more stable} estimates of the error as discussed in Section\nobreakspace \ref {sec:adapt_mlmc} to improve the \nt{overall stability} of MLMC.
The bottom right plot in Figure\nobreakspace \ref {fig:cva_level-plots} shows the ratio \(R_{\ell_0}\) defined in Section\nobreakspace \ref {sec:practical_mlmc} against \(\ell_0\). To ensure MLMC performs optimally, one should start from a level \(\ell_0\) for which \(R_{\ell_0} < 1\). 

In Figure\nobreakspace \ref {fig:cva_complexity}, we see a reduction in cost by a factor of around 7 times when using adaptive opposed \ntg{to} non-adaptive MLMC when computing \(\pll\) to 2 significant figures. We also see that our results closely match the complexities \(\Order{\tol^{-5/2}} \) \ntg{for} non-adaptive sampling and \(\Order{\tol^{-2}\p{\log \tol}^2}\) for adaptive sampling discussed in Section\nobreakspace \ref {sec:adapt_mlmc}, where the factor \(\p{\log\tol}^2=\order{\tol^{-\delta}}\) for any \(\delta>0\) is used to model the limiting convergence rate from Theorem\nobreakspace \ref {thm:ad_mlmc_hierarchy} as \(p\to\infty\). A reference line shows the theoretical \(\Order{\tol^{-5}}\) complexity discussed in Section\nobreakspace \ref {sec:cva_hierarchical_mc} for single-level nested Monte Carlo in Section\nobreakspace \ref {sec:cva_mlmc}. This illustrates savings of several orders of magnitude at the smallest computed error tolerance by using the \nt{adaptive hierarchical estimator} compared to using standard Monte Carlo.

\begin{figure}
	\centering 
	\begin{tabular}{|ccc|}
		\hline
		\ref*{pl:ad} Adaptive (\(\refp = 1.95\)) & \ref*{pl:gam1} $\gamma = 1$ & \ref*{pl:gam2} $\gamma = 2$\\
		\hline
	\end{tabular}
\begin{tabular}{c@{\hskip 2cm}c}
	\begin{tikzpicture}[trim axis left, trim axis right]
	\begin{axis}[
	xlabel = $\ell$,
	ylabel = $C_\ell$,
	grid = both,
	ymode = log,
	ymin = 10^(2), ymax = 10^(7),
	]
	\addplot+[mark = diamond, mark size = 3pt, black, opacity = \opac,] 
	table[x = level, y expr = \thisrow{cost}/\thisrow{M}, col sep = comma,]{data/cva/st-gam1-exp-pi-levels.csv};
	\label{pl:gam1}
	
	\addplot+[mark = o, mark size = 3pt, black, opacity = \opac,] 
	table[x = level, y expr = \thisrow{cost}/\thisrow{M}, col sep = comma,]{data/cva/st-gam2-exp-pi-levels.csv};
	\label{pl:gam2}
	
	\addplot+[mark = asterisk, mark size = 3pt, black, opacity = \opac,] 
	table[x = level, y expr = \thisrow{cost}/\thisrow{M}, col sep = comma,]{data/cva/ad-umlmc-exp-pi-levels.csv};
	\label{pl:ad}
	\end{axis}
	\end{tikzpicture}
	&
	\begin{tikzpicture}[trim axis left, trim axis right]
	\begin{axis}[
	xlabel = $\ell$,
	ylabel = $V_\ell$,
	grid = both,
	ymode = log,
	ymax = 10^(-0.5), ymin = 10^(-5.5),
	]
	\addplot+[mark = diamond, mark size = 3pt, black, opacity = \opac] 
	table[x = level, y expr = \thisrow{V}, col sep = comma,]{data/cva/st-gam1-exp-pi-levels.csv};
	
	\addplot+[mark = o, mark size = 3pt, black, opacity = \opac,] 
	table[x = level, y expr = \thisrow{V}, col sep = comma,]{data/cva/st-gam2-exp-pi-levels.csv};
	
	\addplot+[mark = asterisk, mark size = 3pt, black, opacity = \opac,] 
	table[x = level, y expr = \thisrow{V}, col sep = comma,]{data/cva/ad-umlmc-exp-pi-levels.csv};
	
	\draw(3.8, 0.00002) node[fill = white, shape = rectangle]{\textcolor{gray}{$\Order{2^{-\ell}}$}};
	\addplot+[mark = none, gray, domain = 2:8]{2^(-x)*10^(-2.5)};
	
	\draw(5, 0.04) node[fill = white, shape = rectangle]{\textcolor{gray}{$\Order{2^{-\ell/2}}$}};
	\addplot+[mark = none, densely dashed, gray, domain = 2:8]{2^(-x/2)*10^(-1.5)};
	\end{axis}
	\end{tikzpicture}\\	
	\begin{tikzpicture}[trim axis left, trim axis right]
	\begin{axis}[
	xlabel = $\ell$,
	ylabel = $\kurt{\dhl}$,
	grid = both,
	ymode = log,
	]
	\addplot+[mark = diamond, mark size = 3pt, black, opacity = \opac,] 
	table[x = level, y expr = \thisrow{kurtosis}, col sep = comma,]{data/cva/st-gam1-exp-pi-levels.csv};
	
	\addplot+[mark = o, mark size = 3pt, black, opacity = \opac,] 
	table[x = level, y expr = \thisrow{kurtosis}, col sep = comma,]{data/cva/st-gam2-exp-pi-levels.csv};
	
	\addplot+[mark = asterisk, mark size = 3pt, black, opacity = \opac,] 
	table[x = level, y expr = \thisrow{kurtosis}, col sep = comma,]{data/cva/ad-umlmc-exp-pi-levels.csv};
	\end{axis}
	\end{tikzpicture}
	&
	\begin{tikzpicture}[trim axis left, trim axis right]
	\begin{axis}[
	xlabel = $\ell_0$,
	ylabel = $R_{\ell_0}$,
	grid = both,
	ymin = 0.4, ymax = 1.6,
	]
	\addplot+[mark = diamond, mark size = 3pt, black, opacity = \opac,] 
	table[x = level, y expr = \thisrow{R}, col sep = comma,]{data/cva/st-gam1-exp-pi-ell0.csv};
	
	\addplot+[mark = o, mark size = 3pt, black, opacity = \opac,] 
	table[x = level, y expr = \thisrow{R}, col sep = comma,]{data/cva/st-gam2-exp-pi-ell0.csv};
	
	\addplot+[mark = asterisk, mark size = 3pt, black, opacity = \opac,] 
	table[x = level, y expr = \thisrow{R}, col sep = comma,]{data/cva/ad-umlmc-exp-pi-ell0.csv};
	
	\addplot+[mark = none, black, domain=0:7]{1};
	\end{axis}
	\end{tikzpicture}
\end{tabular}
\caption{Statistics of the correction terms \(\dhl\) plotted against $\ell$ for the MLMC estimators using both adaptive and non-adaptive sampling. From top left to bottom right, the plots show the expected cost, \(C_\ell\), of generating a single sample of \(\dhl\), computed as the number of independent Gaussian and Exponential random variables simulated, the variance \(V_\ell=\var{\dhl}\) and the kurtosis \(\kurt{\dhl}\), each \nt{versus} \(\ell\). The bottom right plot shows the ratio \(R_{\ell_0}\) in Section\nobreakspace \ref {sec:practical_mlmc} \nt{versus} \(\ell_0\), which is used to determine a starting level for the MLMC estimates.}
\label{fig:cva_level-plots}
\end{figure}
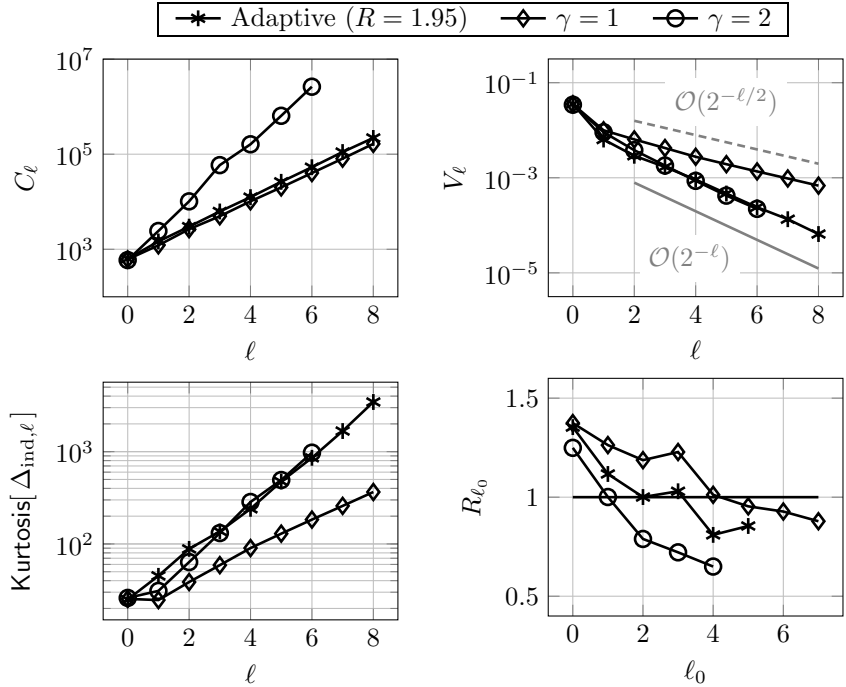 \textsl{}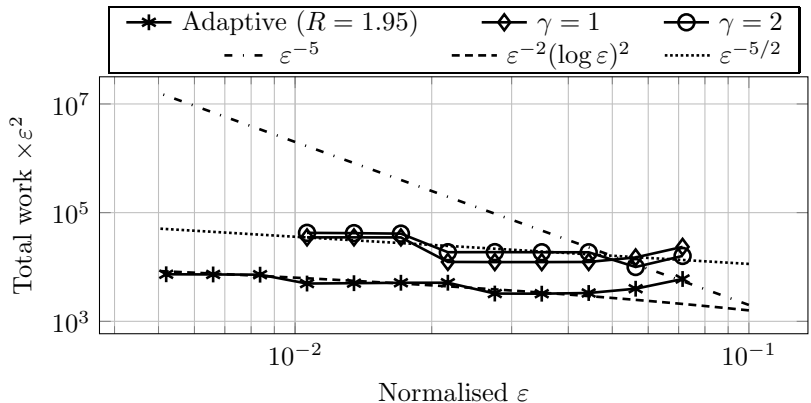
\begin{figure}
	\centering 
	\begin{tabular}{|ccc|}
		\hline
		\ref*{pl:ad} Adaptive (\(\refp = 1.95\)) & \ref*{pl:gam1} $\gamma = 1$ & \ref*{pl:gam2} $\gamma = 2$\\
		\ref*{plot:complexity-MC} \(\tol^{-5}\) & \ref*{pl:complex-log} \(\tol^{-2}\p{\log \tol}^2\) & \ref*{plot:complexity-big} \(\tol^{-5/2}\)\\
		\hline
	\end{tabular}\\	
	\begin{tikzpicture}[trim axis left]
	\begin{axis}[
	xlabel = Normalised $\tol$,
	ylabel = Total work $\times \tol^2$,
	grid = both,
	ymode = log,
	xmode = log,
	ymax = 10^(7.5), width = 11cm, height = 5cm
	]
	\addplot+[mark = diamond, mark size = 3pt, black, opacity = \opac,] 
	table[x expr = \thisrow{tol}/0.035, y expr = \thisrow{cost}*\thisrow{tol}^2, col sep = comma,]{data/cva/st-gam1-exp-pi-mlmc.csv};
	
	\addplot+[mark = o, mark size = 3pt, black, opacity = \opac,] 
	table[x expr = \thisrow{tol}/0.035, y expr = \thisrow{cost}*\thisrow{tol}^2, col sep = comma,]{data/cva/st-gam2-exp-pi-mlmc.csv};
	
	\addplot+[mark = asterisk, mark size = 3pt, black, opacity = \opac,] 
	table[x expr = \thisrow{tol}/0.035, y expr = \thisrow{cost}*\thisrow{tol}^2, col sep = comma,]{data/cva/ad-umlmc-exp-pi-mlmc.csv};

	\addplot+[black, mark = none, loosely dashdotted, domain = 0.1:0.005]{2*x^(-3)};
	\label{plot:complexity-MC}
	\addplot+[black, mark = none, densely dotted, domain = 0.1:0.005]{3600*x^(-0.5)};
	\label{plot:complexity-big}
	\addplot+[black, mark = none, densely dashed, domain = 0.1:0.005]{300*ln(x)^2};
	\label{pl:complex-log}
	\end{axis}
	\end{tikzpicture}
\caption{Total work required to compute the probability of large loss due to changes in the CVA multiplied by $\tol^2$ \nt{versus} $\tol$, normalised by the true value of $\pll$.}
\label{fig:cva_complexity}
\end{figure}  
\section{Conclusion}
Due to large costs induced by the approximation of market factors combined with several nested Monte Carlo averages, accurate estimation of credit risk measures is often {considered} infeasible using traditional Monte Carlo methods. By combining recent developments in MLMC for nested simulation with variance reduction techniques, this paper illustrates an efficient framework for accurate simulation of these risk measures at greatly reduced costs. In the context of the CVA-VaR problem, we are able to reduce the cost of achieving accuracy $\tol$ from $\Order{\tol^{-5}}$ for traditional \nt{nested} Monte Carlo to $\Order{\tol^{-2-\delta}}$, for any \(\delta>0\), using MLMC with an adaptive number of inner samples of the loss. This results in computational savings of several orders of magnitude for realistic problems.  

Future directions of this work include numerical studies for larger portfolios, containing a greater number of counterparties and the posting of collateral as discussed in Sections\nobreakspace \ref {sec:collateralisation} and\nobreakspace  \ref {sec:cva_multiple_counterparties}, and a combination of the methods in this work with multilevel stochastic approximation techniques \cite{Dereich2019,cfl:2023,Frikha2016} to compute the value-at-risk \(\lloss\) associated \ntg{with} a fixed confidence level \(\pll\). Further analysis of the numerical estimator is required to handle non-linear nested market factors such as the initial margin and also to include more general models for default which cannot be simulated \st{exactly}, and will require more advanced importance sampling methods. In more general models for default, an important concept to consider is that of wrong way risk \final{as considered by Glasserman and Yang \cite{glasserman15}, Gregory \cite{Gregory2020} and Hull et al.. \cite{hw12},} where the probability of counterparty default is positively correlated to the risk-neutral valuation of the portfolio, leading to greater credit risk.

\section*{Competing Interests}
The authors declare no competing interests.

\appendix\normalsize
\section{Proofs}\label{app:proofs}
To prove Proposition\nobreakspace \ref {prop:cva_umlmc_moments}, we rely on the following general result to bound moments of antithetic multilevel differences of the form \eqref{eqn:dlantf}.
\begin{lemma}\label{lem:cva_general_antithetic_result}
	Let \(U_{1,\ell-1}^{\p{0}}\) and \(U_{1,\ell-1}^{\p{1}}\) be conditionally independent approximations of \({\nestE{1}}\) at level \(\ell-1\), with finite \(q_1^\textnormal{th}\) moment for some \(q_1\ge 2\). Let \(U_{1,\ell} \defeq \p{U_{1,\ell-1}^{\p{0}} + U_{1,\ell-1}^{\p{1}}}/2\) be an approximation to \({\nestE{1}}\) at level \(\ell\). \nt{For an} \(f:\rset\to\rset\) \nt{satisfying} \MakeUppercase assumption\nobreakspace \ref {assumpt:rec_f_decomposition}, there is a constant \(c> 0\), independent of \(\ell\), such that for all \(2\le q_0\le q_1\) and \(\ell\ge 0\) we have
	\[
	\begin{aligned}
	&\E[\bigg]{\abs[\Big]{f\p{U_{1, \ell}} - \frac{1}{2}\p[\big]{f\p{U_{1, \ell-1}^{\p{0}}} + f\p{U_{1, \ell-1}^{\p{1}}}}}^{q_0}}\\
	 &\le
	 c\bigg( \ntg{\E[\Big]{\abs[\big]{U_{1, \ell-1}^{\p{0}} - {\nestE{1}}}^{q_1}}^{\min\br{2{q_0}/{q_1}, 1}}} \quad + \E[\Big]{\abs[\big]{U_{1,\ell} - {\nestE{1}}}^{q_1}}^{\p{{q_0}+1}/\p{q_1+1}}\bigg).
	\end{aligned}
	\]
\end{lemma}
\ntg{
\begin{text_comment}
	The proof of Lemma\nobreakspace \ref {lem:cva_general_antithetic_result} follows similar steps to \cite[Theorem 4.1]{GilesHajiAli:2018}, \cite[Proposition 5]{Bourgey20}, \cite[Theorem 3]{gg19} and \cite[Theorem 5.2]{giles14antmilstein}. A slightly more general version of this result, which is not specific to unbiased approximations of a nested conditional expectation may be found in \cite[Corollary 2.6]{spence23}.  Since the complexity of MLMC estimates depends on the variance of the multilevel differences, \nt{several existing} results assume \(q_0 = 2\). However, when estimating the triply nested expectation \eqref{eqn:cva_general_nested_prob}, since \MakeUppercase assumption\nobreakspace \ref {assumpt:admlmc} requires more than convergence in the second moment, Proposition\nobreakspace \ref {prop:cva_umlmc_moments} includes the asymptotic behaviour of \(2\le q_0\le q_1\) moments of the multilevel differences. A similar approach for an arbitrary number of nested expectations may be found in \cite[Theorem 1]{zhou2022} and \cite[Theorems 2.2 and 2.4]{syed2023optimal}.
\end{text_comment}
\begin{proof}
	Throughout this proof, consider
	\[
	\dl f \defeq f\p[\bigg]{\frac{U_{1, \ell-1}^{\p{0}} + U_{1, \ell-1}^{\p{1}}}{2}} - \frac{1}{2}\sum_{i=0}^1 f\p{U_{1, \ell-1}^{\p{i}}}.
	\]
	Following the same steps as \cite[Theorem 5.2]{giles14antmilstein}, for some \(\psi > 0\) consider the indicator functions
	\[
	\begin{aligned}
	E_0 &\defeq \I{\inf_{u\in\Gamma}\abs{{\nestE{1}} - u}> \psi}\\
	E_1 &\defeq \I{\abs{U_{1, \ell-1}^{\p{0}} - {\nestE{1}}} \le \psi/2}\\
	E_2 &\defeq \I{\abs{U_{1, \ell-1}^{\p{0}} - U_{1, \ell-1}^{\p{1}}} \le\psi / 2} 
	\end{aligned}
	\]
	and let \(E_j^c\defeq 1 - E_j\) for \(j=0,1,2\).
	Then,
	\[
	\begin{aligned} 
	\E[\big]{\abs{\dl f}^{q_0}} &\le \E[\big]{\abs{\dl f}^pE_0E_1E_2} + \sum_{j=0}^2\E[\big]{\abs{\dl f}^pE_j^c}.
	\end{aligned}  
	\]
	For the first term, note that \(E_0E_1E_2 = 1\) implies that \(f\) is twice differentiable on an open interval of radius \(\psi\) centred on \({\nestE{1}}\) and containing both \(U_{1, \ell-1}^{\p{0}}\) and \(U_{1, \ell-1}^{\p{1}}\). In particular, since \(U_{1,\ell} = \p{U_{1, \ell-1}^{\p{0}} + U_{1, \ell-1}^{\p{1}}}/2\) lies in this interval, and since \(\dl f =0\) on a linearisation of \(f\),  one can use the Lipschitz property of \(f\) and a second order Taylor expansion to conclude that when \(E_0E_1E_2 = 1\), the following two inequalities hold
	\[
	\begin{aligned}
	\abs{\dl f} &\le L_f\abs{U_{1, \ell-1}^{\p{0}} - U_{1, \ell-1}^{\p{1}}}\\
	\abs{\dl f} &\le L_f^\prime\abs{U_{1, \ell-1}^{\p{0}} - U_{1, \ell-1}^{\p{1}}}^2.
	\end{aligned} 
	\]
	\st{Taking} appropriate powers of the above expansions, with \(v = \max\br{0, \p{2q_0-{q_1}}/q_0}\), for \(q_0 \le {q_1}\) it follows that
	\[
	\begin{aligned} 
	\E[\big]{\abs{\Delta_\ell f}^{q_0}E_0E_1E_2} &\le 
	2^{{q_1}-1}\p{L_f}^{v{q_0}}\p{L_f^\prime}^{\p{1-v}{q_0}}\E[\Big]{\abs[\big]{U_{1, \ell-1}^{\p{0}} - {\nestE{1}}}^{\min\br{2{q_0}, {q_1}}}}\\
	&\le 2^{{q_1}-1}\p{L_f}^{v{q_0}}\p{L_f^\prime}^{\p{1-v}{q_0}}\E[\Big]{\abs[\big]{U_{1, \ell-1}^{\p{0}} - {\nestE{1}}}^{q_1}}^{\min\br{2{q_0}/{q_1}, 1}},
	\end{aligned}
	\]
	using H\"older's inequality in the final line.
	
	For the remaining terms, since \(f\) is \(L_f\)-Lipschitz, it follows for \(j=0,1,2\) that
	\[
	\begin{aligned}
	\E[\big]{\abs{\dl f}^{q_0}E_j^c} &\le \p{L_f}^{q_0} \E[\big]{\abs{U_{1, \ell-1}^{\p{0}} - U_{1, \ell-1}^{\p{1}}}^{q_0} E_j^c}\\
	&\le \p{L_f}^{q_0}2^{{q_0}-1}\E[\Big]{\abs[\big]{U_{1, \ell-1}^{\p{0}} - {\nestE{1}}}^{q_1}}^{{q_0}/{q_1}} \E{E_j^c}^{\p{{q_1}-{q_0}}/{q_1}},
	\end{aligned}	
	\]
	using H\"older and Jensen's inequalities in the final line for \({q_0}\le {q_1}\). For \(j = 0\), we have
	\[
	\begin{aligned}
	\E{E_0^c} &= \prob[\bigg]{\inf_{u\in\Gamma}\abs{{\nestE{1}} - u}\le \psi} \le \bar\rho\psi 
	\end{aligned}
	\]
	for any \(\psi<\bar\delta\) under \MakeUppercase assumption\nobreakspace \ref {assumpt:rec_f_decomposition}. For \(j = 1\) we have,
	\[
	\begin{aligned}
	\E{E_1^c} &= \E[\Big]{\I{\abs{U_{1, \ell-1}^{\p{0}} - {\nestE{1}}} > \psi/2}}\\
	&\le 2^{{q_1}}\psi^{-{q_1}}\E[\Big]{\abs[\big]{U_{1, \ell-1}^{\p{0}} - {\nestE{1}}}^{q_1}}.
	\end{aligned}
	\]
	Similar steps show that 
	\[
	\begin{aligned}
	\E{E_2^c} &\le 2^{{q_1}}\psi^{-{q_1}}\E[\big]{\abs{U_{1, \ell-1}^{\p{0}} - U_{1, \ell-1}^{\p{1}}}^{q_1}}\\
	&\le 2^{{q_1}}2^{{q_1}-1}\psi^{-{q_1}}\E[\Big]{\abs[\big]{U_{1, \ell-1}^{\p{0}} - {\nestE{1}}}^{q_1}}
	\end{aligned}
	\]	
	using Jensen's inequality in the final line. To make the above terms of the same order, one should take 
	\[
	\psi = \min\br[\bigg]{1, \frac{\bar\delta}{\sup_{\ell\ge 0}\abs{U_{1, \ell-1}^{\p{0}} - {\nestE{1}}}}} \E[\Big]{\abs[\big]{U_{1, \ell-1}^{\p{0}} - {\nestE{1}}}^{q_1}}^{1/\p{{q_1}+1}}.
	\]
	Hence, there is \(b>0\) independent of \(\ell\) such that for \(j = 0,1,2\)
	\[
	\begin{aligned}
	\E[\big]{\abs{\dl f}^pE_j^c} &\le b\E[\Big]{\abs[\big]{U_{1, \ell-1}^{\p{0}} - {\nestE{1}}}^{q_1}}^{{q_0}/{q_1}}\E[\Big]{\abs[\big]{U_{1, \ell-1}^{\p{0}} - {\nestE{1}}}^{q_1}}^{\p{{q_1}-{q_0}}/{q_1}\p{{q_1}+1}}\\
	&=b \E[\Big]{\abs[\big]{U_{1, \ell-1}^{\p{0}} - {\nestE{1}}}^{q_1}}^{\p{{q_0}+1}/\p{{q_1}+1}},
	\end{aligned}
	\]
	which completes the result.\qed
\end{proof}
}
\begin{proof}[of Proposition\nobreakspace \ref {prop:cva_umlmc_moments}]
	\nt{Throughout this proof, we denote by \(c>0\) an arbitrary constant which is independent of \(\ell\) and \(k\).} We begin by proving that \(\dpz\) has finite expected sampling cost. First, by \MakeUppercase assumption\nobreakspace \ref {assumpt:cva_general_rv_convergence}, we know that the cost of sampling \(\dl[k]\nvar{x}\) is bounded by \(c2^k\). Therefore
	\[
		\begin{aligned}
			\E[\Big]{\text{Cost}\p[\big]{\dl[\rlev{1}]\nvar{x}}} &= \sum_{k=0}^\infty \text{Cost}\p[\big]{\dl[k]\nvar{x}}p_{1, k}\\
			&\nt{\le c}\sum_{k=0}^\infty 2^{\p{1-\zeta_1}k},
		\end{aligned} 
	\] 
	which is finite \nt{using the geometric decay of \(p_{1, k}\)} since \(1<\zeta_1 = 3/2\) by hypothesis. 
	Hence, 
	\[
		\begin{aligned}
			\E[\Big]{\textnormal{Cost}\p[\big]{\appEub{1}{\ell}\p{\nvar[\ell]{y}} }} &\propto 2^\ell,
		\end{aligned}
	\]
	from which it follows from \MakeUppercase assumption\nobreakspace \ref {assumpt:cva_general_rv_convergence} and similar steps as above that
	\[
		\begin{aligned} 
		\E[\Big]{\textnormal{Cost}\p[\big]{\dpz}} &\nt{\le c}\sum_{\ell=0}^\infty 2^{\p{1-\zeta_0}\ell}, 
		\end{aligned} 
	\]
	which is finite since it is assumed that \(\zeta_0 = 17/16 > 1\).\\
	
	It remains to provide bounds on the moments of \(\dpz\). We will show, for the random variable \(\dl[\ell]^{\p{\textnormal{UB}}} f\) defined in \eqref{eqn:dlantf}, that for fixed \(\ell\ge 1\) and \(2\le q_0<3\), we have
	\begin{equation}\label{eqn:dl_ub_moment_bound}
		\E[\Big]{\abs[\big]{\dl[\ell]^{\p{\textnormal{UB}}} f}^{q_0}} \nt{\le c} {2^{-3\p{q_0+1}\ell/8 + \delta\ell}},
	\end{equation}
	for any \(\delta>0\). We split a proof of \eqref{eqn:dl_ub_moment_bound} into three steps:
	\begin{enumerate}
		\item We first show that, for \(2 \le q_1<3\),
		\begin{equation}\label{eqn:proof_1}
		\E[\Big]{\abs[\big]{\appEub{1}{\ell}\p{\nvar[\ell]{y}} - {\nestE{1}}}^{q_1}} \nt{\le c} {2^{-q_1\ell/2}}.
		\end{equation}
		\item Secondly,for \(2\le q_0\le q_1\), we prove
		\begin{equation}\label{eqn:proof_2}
			\E[\Big]{\abs[\big]{\appEub{1}{\ell}\p{\nvar[\ell]{y}} -\appEub{1}{\ell}\p{\nvar[\ell-1]{y}}}^{q_0}} \nt{\le c} {2^{{-3\ell/2}}}.
		\end{equation}
		\item Use Lemma\nobreakspace \ref {lem:cva_general_antithetic_result} to bound moments of \(\dl[\ell]^{\p{\textnormal{UB}}} f\). 
	\end{enumerate}
	
	\textbf{Step 1.} For \(q_1< 3\), it follows from the definition of \(\rlev{1}\) with \(\zeta_1=3/2\) that
	\[
		\begin{aligned}
			\E[\Big]{\abs[\big]{\dl[\rlev{1}]\nvarcond[none][\ell]{x}{y} p_{1,\rlev{1}}^{-1}}^{q_1}} &= \sum_{k=0}^\infty p_{1,k}^{1-q_1}\E[\Big]{\abs[\big]{ \dl[k]\nvarcond[none][\ell]{x}{y}}^{q_1}}\\
			&\le p_{1,0}^{1-q_1}c\sum_{k=0}^{\infty} 2^{3\p{q_1-1}k/2}2^{-q_1 k}\\
			&<\infty.
		\end{aligned}
	\]
	Therefore, \ntg{using \MakeUppercase assumption\nobreakspace \ref {assumpt:cva_general_rv_convergence}, Jensen's inequality, the tower property and the discrete Burkholder-Davis-Gundy inequality following similar steps to \cite[Lemma 2.5]{GilesHajiAli:2018},} we obtain the bound
	\[
	\begin{aligned}
	&\E[\Big]{\abs[\big]{\appEub{1}{\ell}\p{\nvar[\ell]{y}} - {\nestE{1}}}^{q_1}}\\
	&\le 2^{q_1 - 1}\p[\bigg]{\E[\Big]{\abs[\big]{\appEub{1}{\ell}\p{\nvar[\ell]{y}} - {\nestE{1}}^{\br{\nvar[\ell]{y}}}}^{q_1}} + \E[\Big]{\abs[\big]{{\nestE{1}}^{\br{\nvar[\ell]{y}}} - {\nestE{1}}}^{q_1}}} \\
	&\nt{\le c} {2^{-q_1\ell/2}}.
	\end{aligned}
	\]
	
	\textbf{Step 2.} For \(\ell \ge 1\), denote the double difference 
	\[
		\begin{aligned}
			\dl \dl[k]\nvar{x} &\defeq \dl[k]\nvarcond[none][\ell]{x}{y} - \dl[k]\nvarcond[none][\ell-1]{x}{y}\\
			&= \nvarcond[k][\ell]{x}{y} - \nvarcond[k-1][\ell]{x}{y} - \nvarcond[k][\ell]{x}{y} + \nvarcond[k-1][\ell]{x}{y}
		\end{aligned}
	\] 
	and let 
	\[
		\overline{\dl\dl[\rlev{1}]\nvar{x}} \defeq  \dl[\rlev{1}]\nvar{x} p_{1,\rlev{1}}^{-1} - \E[\Big]{\dl \dl[\rlev{1}]\nvar{x} p_{1,\rlev{1}}^{-1}\given \nvar[\ell]{y},\nvar[\ell-1]{y}}
	\] 
	represent the centralised random double differences with mean zero. Under \MakeUppercase assumption\nobreakspace \ref {assumpt:cva_general_rv_convergence} it follows that
	\begin{equation}\label{eqn:cva_double_diff_bound}
		\E[\big]{\abs{\dl \dl[k]\nvar{x}}^{q_0}} \nt{\le c}2^{-q_0\max\br{\ell,k}}.
	\end{equation}
	\ntg{Using the triangle and discrete Burkholder-Davis-Gundy inequalities, since \st{the random variable} \(\overline{\dl\dl[\rlev{1}]\nvar{x}}\) has mean zero, \st{and} following similar steps to \cite[Lemma 2.5]{GilesHajiAli:2018},} we have 
	\[
		\begin{aligned}
			&\E[\Big]{\abs[\big]{\appEub{1}{\ell}\p{\nvar[\ell]{y}} -\appEub{1}{\ell}\p{\nvar[\ell-1]{y}}}^{q_0}}\\ &\le\E[\bigg]{\abs[\bigg]{\frac{1}{N_{1,\ell}} \sum_{n=1}^{N_{1,\ell}} \dl \dl[\rlev{1}^{\p{n}}]\nvarcond[none][none][n]{x}{y}}^{q_0}}\\
			&\le 2^{q_0-1}\p[\bigg]{\E[\bigg]{\abs[\bigg]{\frac{1}{N_{1,\ell}} \sum_{n=1}^{N_{1,\ell}} \overline{\dl \dl[\rlev{1}^{\p{n}}]\nvarcond[none][none][n]{x}{y}}}^{q_0}} + \E[\bigg]{\abs[\Big]{\E[\big]{\dl \dl[\rlev{1}]\nvar{x} p_{1,\rlev{1}}^{-1}\given \nvar[\ell]{y},\nvar[\ell-1]{y}}}^{q_0}}}\\
			&\nt{\le c}_{q_0}\p[\bigg]{N_{1,\ell}^{-q_0/2}\E[\Big]{\abs[\big]{\dl\dl[\rlev{1}]\nvar{x} p_{1,\rlev{1}}^{-1}}^{q_0}} + \E[\bigg]{\abs[\Big]{\E[\big]{\dl \dl[\rlev{1}]\nvar{x} p_{1,\rlev{1}}^{-1}\given \nvar[\ell]{y},\nvar[\ell-1]{y}}}^{q_0}}},
		\end{aligned}
	\] 
	where the constant \(c_{q_0}\) depends only on \(q_0\).
	The latter term can be bounded using \eqref{eqn:cva_ub_innermost}, Jensen's inequality and the tower property to obtain
	\[
		\begin{aligned}
			\E[\bigg]{\abs[\Big]{\E[\big]{\dl \dl[\rlev{1}]\nvar{x} p_{1,\rlev{1}}^{-1}\given \nvar[\ell]{y},\nvar[\ell-1]{y}}}^{q_0}} &= \E[\bigg]{\abs[\Big]{\E[\big]{\nvarcond[none][\ell]{x}{y} - \nvarcond[none][\ell-1]{x}{y}\given\nvar[\ell]{y},\nvar[\ell-1]{y}}}^{q_0}}\\
			&\le \E[\Big]{\abs[\big]{\nvarcond[none][\ell]{x}{y} - \nvarcond[none][\ell-1]{x}{y}}^{q_0}}\\
			&\nt{\le c}2^{-q_0\ell},
		\end{aligned}
	\]
	where we used \MakeUppercase assumption\nobreakspace \ref {assumpt:cva_general_rv_convergence} in the final line. For the remaining term, by the definition of \(\rlev{1}\) with \(\zeta_1 = 3/2\) and using \eqref{eqn:cva_double_diff_bound}, we have
	\[
		\begin{aligned}
			\E[\Big]{{\abs[\big]{\dl\dl[\rlev{1}]\nvar{x} p_{1,\rlev{1}}^{-1}}^{q_0}}} &= \sum_{k=0}^\infty p_{1,k}^{1-q_0}\E[\Big]{\abs[\big]{\dl \dl[k]\nvar{x}}^{q_0}}\\
			&\le p_{1,0}^{1-q_0}c\sum_{k=0}^{\infty} 2^{3\p{q_0-1}k/2}2^{-q_0\max\br{\ell, k}}\\
			&\nt{\le c} {2^{\p{q_0 - 3}\ell/2}},
		\end{aligned}
	\]
	where we use the fact that \(q_0 < 3\) so that the geometric series is convergent.
	
	The previous results, with \(N_{1,\ell}\propto 2^\ell\) are enough to show that 
	\[
	\begin{aligned}
		\E[\Big]{\abs[\big]{\appEub{1}{\ell}\p{\nvar[\ell]{y}} -\appEub{1}{\ell}\p{\nvar[\ell-1]{y}}}^{q_0}} &\nt{\le c} {2^{-3\ell/2} + 2^{-q_0\ell}}\\
		&\nt{\le c} {2^{-3\ell/2}},
	\end{aligned}
	\] 
	where the last line follows since \(q_0\ge 2\). \\
	
	\textbf{Step 3.} Using the fact that \(f\) satisfies \MakeUppercase assumption\nobreakspace \ref {assumpt:rec_f_decomposition}, it follows from Lemma\nobreakspace \ref {lem:cva_general_antithetic_result}, \eqref{eqn:proof_1} and \eqref{eqn:proof_2} that for \(\ell\ge 1\) and \(2\le q_0\le q_1\) we have 
	\[
		\begin{aligned}
		\E[\Big]{\abs[\big]{\dl[\ell]^{\p{\textnormal{UB}}} f}^{q_0}} &\nt{\le c} \ntg{\E[\Big]{\abs[\big]{\appEub{1}{\ell}\p{\nvar[\ell]{y}} -\appEub{1}{\ell}\p{\nvar[\ell-1]{y}}}^{q_1}}^{\min\br{2q_0/q_1,1}}}\\
		&\quad + \ntg{c}\E[\Big]{\abs[\big]{\appEub{1}{\ell}\p{\nvar[\ell]{y}} - {\nestE{1}}}^{q_1}}^{\p{q_0+1}/\p{q_1+1}}\\
		&\nt{\le c} {\p[\Big]{\ntg{2^{-\min\br{2q_0, q_1}\ell/2}} + 2^{-q_1\p{q_0+1}\ell/2\p{q_1 + 1}}}}\\
		&\nt{\le c} {2^{-3\p{q_0 + 1}\ell/8 + \delta\ell}},
		\end{aligned}
	\]
	for any \(\delta>0\), where the last line follows from taking \(q_1\) \ntg{sufficiently close to 3},
	completing the desired bound on \(\E{\abs{\dl[\ell]^{\p{\textnormal{UB}}} f}^{q_0}}\).
	
	\nt{It} follows from the definition of \(\rlev{0}\), with \(\zeta_0 = 17/16\), that
	\[
		\E[\Big]{\abs[\big]{\dpz}^{q_0}} \nt{\le c} \sum_{\ell=0}^\infty 2^{17\p{q_0 - 1}\ell/16}2^{-3\p{q_0 + 1}\ell/8 + \delta\ell},
	\]
	which is finite provided
	\[
	\begin{aligned}
	\frac{17}{16}\p{q_0 - 1} &< \frac{3\p{q_0 + 1} - \delta}{8}.
	\end{aligned}
	\]
	By taking \(\delta>0\) small enough, the above relation holds for any \(2\le q_0 < 23/11\), completing the proof. \qed
\end{proof}

\begin{proof}[of Theorem\nobreakspace \ref {thm:ad_mlmc_hierarchy}]
	In \cite[Proposition 3.1 and Lemma 3.2]{hajiali2021adaptive} it is shown that under \MakeUppercase assumption\nobreakspace \ref {assumpt:admlmc} for \(\dhl\) defined by \eqref{eqn:dhl_adapt}, with refinement parameter \(1<\refp<2\), we have \({\E{\dhl^2} = \Order{N_{0,\ell}^{-p/\p{p+2}}}}\) and \(\E{N_{0, \ell+\eta_\ell}} = \Order{N_{0,\ell}}\), with the former result implying the bound on \(V_\ell=\var{\dhl}\). Furthermore, \(\E{\text{Cost}\p{\dhl}} = N_{0,\ell}\), hence the total computational cost \nt{satisfies} \eqref{eqn:mlmc_cost}. Substituting the asymptotic form for \(V_\ell\), we get 
	\(
	\text{Cost}\p{\pll^{\p{\textnormal{MLMC}}}_{\Mvec^\star, L};\tol} = \Order{\tol^{-2}\p{\sum_{\ell=0}^L N_{0,\ell}^{1/\p{p+2}}}^2}
	\).
	If \(\abs{\E{\dhl}} = \Order{N_{0,\ell}^{-1}}\) then it suffices from standard MLMC complexity theory to take \(L\) \ntg{such that \(N_{0,L}^{-1}=\Order{\tol}\),} providing the desired bound. In the absence of such a bound, one can use \(\abs{\E{\dhl}}\le \E{\dhl^2}^{1/2} = \Order{N_{0,\ell}^{-p/2\p{p+2}}}\). In this case, \ntg{one should take \(L\) such that \(N_{0,L}^{-1}=\Order{\tol^{p/2\p{p+2}}}\),} which provides the alternative bound on the cost.\qed
\end{proof} \section{Bayesian Estimates of the Error}\label{app:bayes_err}
In this appendix, we describe an approach for obtaining more robust estimates of the error \nt{and variance estimate in} MLMC \eqref{eqn:cva_pll_mlmc}. We describe a Bayesian approach, based on a similar method in \cite{Elfverson:2016selectiverefinement}. Throughout this section, let \(V_\ell = \var{\dhl}\) and \(E_\ell=\abs{\E{\dhl}}\), where \(\dhl\) represent either the non-adaptive multilevel differences defined through \eqref{eqn:dhl} or the adaptive differences \eqref{eqn:dhl_adapt}. For  \(\ell\ge 0\) there exists \(p_\ell, q_\ell\in\sq{0,1}\) such that the multilevel correction terms \(\dhl\) take the form
\[
\dhl = 
\begin{cases}
1 & \text{probability } p_\ell\\
-1 & \text{probability } q_\ell\\
0 & \text{probability } 1 - p_\ell - q_\ell.
\end{cases}
\]
In particular,
\[
\begin{aligned}
V_\ell \le p_\ell + q_\ell, \quad E_\ell = \abs{p_\ell - q_\ell}.
\end{aligned}
\]
When \(\ell\) is large, Theorem\nobreakspace \ref {thm:ad_mlmc_hierarchy} implies that both \(p_\ell, q_\ell \ll 1\). Heuristically, when the number of samples \(M_\ell\) of \(\dhl\) is small, it is extremely unlikely we will observe a non-zero sample, creating unstable approximations to \(V_\ell\) and \(E_\ell\). The idea is to impose a Bayesian prior on \(p_\ell\) and  \(q_\ell\) which incorporates knowledge of the theoretical convergence rates of \(V_\ell\) and \(E_\ell\) such that when the number of samples \(M_\ell\) is low, we may rely more heavily on the theoretical rates than on the observations, and vice-versa as \(M_\ell\) becomes large.

Specifically, assume that we know a relationship of the form 
\(
V_\ell \approx a_0 2^{-\beta\ell},
\)
where the constant \(\beta\) can be obtained \nt{a-priori given} Theorem\nobreakspace \ref {thm:ad_mlmc_hierarchy} or estimated from data and \(a_0\) may be reasonably estimated using information from coarse levels where samples of \(\dhl \) are inexpensive. We impose a Dirichlet prior density on the quantities \(p_\ell, q_\ell\)
\[
\dirprior\p{p, q} = \p{1 - p - q}^{c_\ell - 1}p^\kbe q^\kbe,
\]
for constants \(c_\ell, \kbe > 0\). For the robustness of the MLMC estimator, it is preferable to over-estimate opposed to under-estimate \(V_\ell\). Therefore, we require the condition  
\(
\E_{\dirprior}{p_\ell + q_\ell} \ge V_\ell,
\) 
which is true provided 
\(
c_\ell \le 2\p{\kbe+1}\p{V_\ell^{-1} - 1}.
\)
Hence, using \(V_\ell \approx a_0 2^{-\beta\ell}\), a reasonable conservative estimate is
\(
c_\ell = 2\kbe\p{a_0^{-1}2^{\beta\ell} - 1}.
\)
Given observations \(\{\dhl[m]\}_{m=1}^{M_\ell}\), define 
\[
\nu_\ell^{\pm} \defeq \sum_{m=1}^{M_\ell} \I{\dhl^{\p{m}}= \pm 1}.
\]
Then, the maximum-a-posteriori estimate of \(p_\ell + q_\ell\) is 
\[
\hat V_\ell \defeq \frac{\nu_\ell^+ + \nu_\ell^- + 2\kbe}{M_\ell + 2\kbe a_0^{-1}2^{\beta\ell} - 1}.
\]
{The parameter \(\kbe\) can be chosen to control how confident we {are} that \(V_\ell \le \hat V_\ell\) for a small number of samples \(M_\ell\). Larger values of \(\kbe\) produce more reliable{, conservative} bounds.}

To estimate \(E_\ell = \abs{p_\ell-q_\ell} \) we construct biased estimates for \(p_\ell - q_\ell\) and \(q_\ell - p_\ell\) separately. First, in the case where \(p_\ell > q_\ell\), we wish to estimate of \(p_\ell - q_\ell\). In doing so, we impose a \(\text{Dirichlet}\p{d_\ell, \jbe + 1, 1}\) prior on \(p_\ell, q_\ell\) with density 
\[
\overline{\dirprior}\p{p, q} = \p{1 - p - q}^{d_\ell - 1}p^\jbe,
\]
for constants \(d_\ell\) and \(\jbe\).
Given the observations \(\{\dhl^{\p{m}}\}_{m=1}^{M_\ell}\) the posterior distribution becomes a \(\text{Dirichlet}\p{d_\ell, \nu_\ell^+ + \jbe +1, \nu_\ell^- + 1}\) distribution with maximum a posteriori estimator given by 
\[
\frac{\nu_\ell^+ - \nu_\ell^- + \jbe}{M_\ell + \jbe + d_\ell - 1}.
\]
Under the prior distribution we have 
\begin{equation}\label{eqn:bias_p_q}
\E_{\overline{\dirprior}}{p_\ell - q_\ell} = \frac{\jbe}{\jbe + d_\ell + 2},
\end{equation}
thus enforcing the condition 
\[
\E_{\overline{\dirprior}}{p_\ell - q_\ell} \ge e_02^{-\alpha\ell} \approx \abs{p_\ell - q_\ell}
\]
requires
\begin{equation}\label{eqn:dl_bd}
d_\ell \le \jbe\p{e_0^{-1}2^{\alpha\ell} - 1} - 2.
\end{equation}
In the case where \(q_\ell > p_\ell\) one can repeat this process by interchanging the roles of \(p_\ell\) and \(q_\ell\) in \eqref{eqn:bias_p_q}. Taking the maximum of the two resulting estimators, with equality in \eqref{eqn:dl_bd}, \nt{yields}
\[
\hat E_\ell \defeq \frac{\abs{\nu_\ell^+ - \nu_\ell^-} + \jbe}{M_\ell + \jbe e_0^{-1}2^{\alpha\ell} - 3}.
\]
\nt{This method is used for the numerical experiments in Section\nobreakspace \ref {sec:cva_num}, where we set the constants \(\kbe = \jbe = 1\) therein.}

\bibliographystyle{spmpsci}
\bibliography{references}
\end{document}